\documentclass[longauth]{aa} 

\usepackage[breaklinks, colorlinks, citecolor=blue, linkcolor=blue]{hyperref}
\usepackage{graphicx}
\usepackage{txfonts}

\makeatletter
\DeclareRobustCommand{\ion}[2]{%
  \text{#1\,\check@mathfonts\fontsize\sf@size\z@\selectfont #2}%
}
\makeatother

\newcommand{\hii}{\ion{H}{II}}
\newcommand{\nii}{\ion{N}{II}}
\newcommand{\sii}{\ion{S}{II}}
\newcommand{\oiii}{\ion{O}{III}}
\begin{document}

   \title{Star formation scaling relations at ${\sim}100$~pc from PHANGS: Impact of completeness and spatial scale}
   \titlerunning{Resolved scaling relations with PHANGS}


   \author{
   I. Pessa\inst{\ref{mpia}} \and 
   E. Schinnerer\inst{\ref{mpia}} \and 
   F. Belfiore\inst{\ref{inaf}} \and 
   E. Emsellem\inst{\ref{eso}}  \and 
   A.~K.~Leroy\inst{\ref{ohio}} \and 
   A. Schruba\inst{\ref{mpe}} \and
   J.~M.~D.~Kruijssen\inst{\ref{rechen}} \and
   H.-A. Pan\inst{\ref{mpia}} \and 
   G. A. Blanc\inst{\ref{uch},\ref{carn}} \and 
   P. Sanchez-Blazquez\inst{\ref{ucm}} \and 
   F. Bigiel\inst{\ref{bonn}} \and 
   M. Chevance\inst{\ref{rechen}} \and 
   E. Congiu\inst{\ref{uch}} \and
   D. Dale\inst{\ref{wyo}} \and 
   C. M. Faesi\inst{\ref{Massa}} \and
   S.~C.~O. Glover\inst{\ref{zah}} \and 
   K. Grasha\inst{\ref{Canb}} \and 
   B. Groves\inst{\ref{Canb}} \and 
   I. Ho\inst{\ref{mpia}} \and 
   M. Jiménez-Donaire\inst{\ref{oan},\ref{hsc}} \and 
   R. Klessen\inst{\ref{zah},\ref{zw}} \and 
   K. Kreckel\inst{\ref{rechen}} \and
   E. W. Koch\inst{\ref{cfa}} \and
   D. Liu\inst{\ref{mpia}} \and 
   S. Meidt\inst{\ref{Gent}} \and
   J. Pety\inst{\ref{iram}} \and
   M. Querejeta\inst{\ref{oan}} \and 
   E. Rosolowsky\inst{\ref{alb}} \and
   T. Saito\inst{\ref{mpia}} \and 
   F. Santoro\inst{\ref{mpia}} \and 
   J. Sun\inst{\ref{ohio}} \and 
   A. Usero\inst{\ref{oan}} \and
   E. J. Watkins\inst{\ref{rechen}} \and
   T.~G.~Williams\inst{\ref{mpia}}
   }
   \institute{Max-Planck-Institute for Astronomy, K\"onigstuhl 17, D-69117 Heidelberg, Germany\label{mpia}\\
        \email{pessa@mpia.de}
        \and INAF — Osservatorio Astrofisico di Arcetri, Largo E. Fermi 5, I-50125, Florence, Italy\label{inaf}
        \and European Southern Observatory, Karl-Schwarzschild-Stra{\ss}e 2, 85748 Garching, Germany\label{eso}
        \and Department of Astronomy, The Ohio State University, 140 West 18th Avenue, Columbus, OH 43210, USA\label{ohio}
        \and Max-Planck-Institute for extraterrestrial Physics, Giessenbachstra{\ss}e 1, D-85748 Garching, Germany\label{mpe}
        \and Astronomisches Rechen-Institut, Zentrum f\"ur Astronomie der Universit\"at Heidelberg, M\"onchhofstra{\ss}e 12-14, D-69120 Heidelberg, Germany\label{rechen}
        \and Departamento de Astronom\'ia, Universidad de Chile, Santiago,Chile\label{uch}
        \and Observatories of the Carnegie Institution for Science, Pasadena, CA, USA\label{carn}
        \and Departamento de F\'isica de la Tierra y Astrof\'isica, Universidad Complutense de Madrid, E-28040 Madrid, Spain \label{ucm}
        \and Argelander-Institut f\"ur Astronomie, Universit\"at Bonn, Auf dem H\"ugel 71, D-53121 Bonn, Germany\label{bonn}
        \and Department of Physics and Astronomy, University of Wyoming, Laramie, WY 82071, USA\label{wyo}
        \and Dept. of Astronomy, University of Massachusetts - Amherst, 710 N. Pleasant Street, Amherst, MA 01003, USA\label{Massa}
        \and Universit\"at Heidelberg, Zentrum f\"ur Astronomie, Institut f\"ur theoretische Astrophysik, Albert-Ueberle-Stra{\ss}e 2, D-69120, Heidelberg, Germany\label{zah}
        \and Research School of Astronomy and Astrophysics, Australian National University, Canberra, ACT 2611, Australia\label{Canb}
        \and Observatorio Astronómico Nacional (IGN), C/Alfonso XII, 3, E-28014 Madrid, Spain\label{oan}
        \and Centro de Desarrollos Tecnológicos, Observatorio de Yebes (IGN), 19141 Yebes, Guadalajara, Spain\label{hsc}
        \and Universit\"at Heidelberg, Interdisziplin\"ares Zentrum f\"ur Wissenschaftliches Rechnen, Im Neuenheimer Feld 205, D-69120 Heidelberg, Germany\label{zw}
        \and Center for Astrophysics $\mid$ Harvard \& Smithsonian, 60 Garden St., Cambridge, MA 02138, USA\label{cfa}
        \and Sterrenkundig Observatorium, Universiteit Gent, Krijgslaan 281 S9, B-9000 Gent, Belgium\label{Gent}
        \and Institut de Radioastronomie Millim\'etrique (IRAM), 300 Rue de la Piscine, F-38406 Saint Martin d’H`eres, France\label{iram}
        \and Department of Physics, University of Alberta, Edmonton, AB T6G 2E1, Canada\label{alb}
             }

   \date{Received March 8, 2021; Accepted April 18, 2021}

 
  \abstract
   {}
  {The complexity of star formation at the physical scale of molecular clouds is not yet fully understood. We investigate the mechanisms regulating the formation of stars in different environments within nearby star-forming galaxies from the Physics at High Angular resolution in Nearby GalaxieS (PHANGS) sample.}
   {Integral field spectroscopic data and radio-interferometric observations of $18$ galaxies were combined to explore the existence of the resolved star formation main sequence ($\Sigma_\mathrm{stellar}$ versus $\Sigma_\mathrm{SFR}$), resolved Kennicutt--Schmidt relation ($\Sigma_\mathrm{mol. gas}$ versus $\Sigma_\mathrm{SFR}$), and resolved molecular gas main sequence ($\Sigma_\mathrm{stellar}$ versus $\Sigma_\mathrm{mol. gas}$), and we derived their slope and scatter at spatial resolutions from $100$~pc to $1$~kpc (under various assumptions).}
{All three relations were recovered at the highest spatial resolution ($100$~pc). Furthermore, significant variations in these scaling relations were observed across different galactic environments. 
The exclusion of non-detections has a systematic impact on the inferred slope as a function of the spatial scale. Finally, the scatter of the $\Sigma_\mathrm{mol. gas + stellar}$ versus $\Sigma_\mathrm{SFR}$ correlation is smaller than that of the resolved star formation main sequence, but higher than that found for the resolved Kennicutt--Schmidt relation.}
{The resolved molecular gas main sequence has the tightest relation at a spatial scale of $100$~pc (scatter of $0.34$~dex), followed by the resolved Kennicutt--Schmidt relation ($0.41$~dex) and then the resolved star formation main sequence ($0.51$~dex). This is consistent with expectations from the timescales involved in the evolutionary cycle of molecular clouds. Surprisingly, the resolved Kennicutt--Schmidt relation shows the least variation across galaxies and environments, suggesting a tight link between molecular gas and subsequent star formation. The scatter of the three relations decreases at lower spatial resolutions, with the resolved Kennicutt--Schmidt relation being the tightest ($0.27$~dex) at a spatial scale of $1$~kpc.  Variation in the slope of the resolved star formation main sequence among galaxies is partially due to different detection fractions of $\Sigma_\mathrm{SFR}$ with respect to $\Sigma_\mathrm{stellar}$. }
\maketitle
\section{Introduction}
\label{sec:intro}
In the current paradigm of evolution of galaxies, star formation occurs inside cold and dense molecular gas clouds. This process is regulated by complex small- and large-scale physics, such as feedback from recently born stars and supernovae resulting from the death of the most massive stars, magnetic fields or hydrostatic pressure exerted by the baryonic mass \citep{Kennicutt2012, Ostriker2011, Heyer2015, kruijssen2019, Krumholz2019, Chevance2020}. As a result of the interplay of such regulating mechanisms, different scaling relations at galactic scales arise between the total amount of star formation in a galaxy and the physical quantities that contribute to its regulation.

In this regard, the star formation main sequence (SFMS) is a tight (scatter of ${\sim}0.3$ dex) relation between the total star formation rate (SFR) of a galaxy and its total stellar mass. It consists of a power law, with a slope of ${\sim}1$, and it has been studied in the local Universe and at a higher redshift \citep{Brinchmann2004, Daddi2007, Noeske2007, Salim2007, Lin2012, Whitaker2012, Speagle2014, Saintonge2016, Popesso2019}.
Similarly, the Kennicutt--Schmidt relation has been extensively studied as it offers an alternative perspective on what drives the SFR in a galaxy. It correlates the total SFR with the total amount of gas and is consistent with a power law of order unity, even though the methodology does have an impact on the specific quantitative description \citep{Schmidt1959, Kennicutt1998, Wyder2009, Genzel2010, Tacconi2010, Bigiel2011, Schruba2011, Genzel2012}.

Recent studies have demonstrated that these relations hold down to kiloparsec and sub-kiloparsec spatial scales, although their scatter is expected to raise below a critical spatial scale due to statistical undersampling of the star formation process \citep{Schruba2010, Feldmann2011, Kruijssen2014, Kruijssen2018}. The so-called resolved star formation main sequence \citep[rSFMS;][]{CanoDiaz2016, Abdurro2017, Hsieh2017, Liu2018, Medling2018, Lin2019, Ellison2020, Morselli2020} and resolved Kennicutt--Schmidt relation \citep[rKS;][]{Bigiel2008, Leroy2008, Onodera2010, Schruba2011, Ford2013, Leroy2013, Kreckel2018, Williams2018, Dey2019} correlate the local star formation rate surface density ($\Sigma_\mathrm{SFR}$) with the local stellar mass surface density ($\Sigma_\mathrm{stellar}$) and molecular gas surface density ($\Sigma_\mathrm{mol.gas}$), respectively. However, there remains debate about the slope of these scaling relations, and it is known that their values depend on the specific approach used for their calculation \citep[e.g., fitting technique;][]{Calzetti2012, Reyes2019}. Previous works have uncovered that their slope and scatter may link the physics from global to smaller scales. Moreover, the scatter of the rKS has been interpreted as individual star-forming regions undergoing independent evolutionary life cycles \citep{Schruba2010, Feldmann2011}, and its dependence on spatial scale provides insight into the timescales of the star formation cycle \citep{Kruijssen2014, Kruijssen2018}. Additionally,
\citet{Bacchini2019a, Bacchini2020} found a tight correlation between the gas and the SFR volume densities in nearby disk galaxies, suggesting that the scatter of the rKS could be in part related to projection effects caused by the flaring scale height of gas disks. In addition to the rSFMS and the rKS, recently \citet{Lin2019}, \citet{Ellison2020}, and \citet{Morselli2020} reported the existence of a resolved molecular gas main sequence (rMGMS) as the correlation between the local $\Sigma_\mathrm{mol. gas}$ and local $\Sigma_\mathrm{stellar}$, and several other studies have explored different forms of correlations between these locally measured quantities \citep{Matteucci1989, Shi2011, Dib2017, Shi2018, Dey2019, Lin2019, Barrera-Ballesteros2021}

The existence of these scaling relations suggests that their global counterparts are an outcome of the local mechanisms that drive star formation, and studying these relations and their respective scatter provides powerful insight into the local physical processes regulating the formation of stars. While the correlation between $\Sigma_\mathrm{SFR}$ and $\Sigma_\mathrm{mol.gas}$ is physically more intuitive, since molecular gas is the fuel for new stars, the origin of the relation between $\Sigma_\mathrm{SFR}$ and $\Sigma_\mathrm{stellar}$ is more uncertain. It can be understood as the interplay between local hydrostatic pressure of the disk and feedback mechanisms regulating the formation of new stars \citep{Ostriker2010, Ostriker2011}. On the other hand, using the ALMA-MaNGA QUEnch and STar formation (ALMaQUEST) survey \citep{Lin2020}, \citet{Lin2019} conclude that the rSFMS likely arises due to the combination of the rKS and the rMGMS, while the latter results from stars and gas following the same underlying gravitational potential. 

In this paper we aim to probe these scaling relations to understand the mechanisms that locally regulate star formation, using ${\sim}100$~pc spatial resolution data from the PHANGS survey. We measured the slope and scatter of these resolved scaling relations in individual galaxies and in our combined sample. We study the universality of these relations and the processes driving galaxy-to-galaxy variations. Additionally we study how the spatial scale of the data, fitting approach, and systematic assumptions, such as the CO-to-H$_{2}$ conversion factor, impact our results.

This paper is structured as follows: In Sec.~\ref{sec:data} we present our data set, and in Sec.~\ref{sec:methods} we describe the methods used in our analysis. In Sec.~\ref{sec:results} we present our main results. In Sec.~\ref{sec:disc} we discuss our findings and their implications. Finally our conclusions are summarized in Sec~\ref{sec:summary}.

\section{Data}
\label{sec:data}

We use a sample of $18$ star-forming galaxies, all of them are close to the SFMS of galaxies. These galaxies represent a subsample of the Physics at High Angular resolution in Nearby GalaxieS (PHANGS\footnote{\url{http://phangs.org/}}) survey \citep{Leroy2021b}. The galaxies from the PHANGS survey have been selected to have a distance lower than $20$~Mpc to resolve the typical scale of star-forming regions ($50{-}100$~pc) and to be relatively face on ($i<60^{\circ}$) to limit the effect of extinction and make the identification of clouds easier. These galaxies have been chosen to be a representative set of galaxies where most of the star formation is occurring in the local Universe. Our sample is summarized in Table~\ref{tab:sample} where we use the global parameters as reported by \citet{Leroy2021b} based on the distance compilation of \citet{Anand2021} as well as the inclinations determined by \citet{Lang2020}.

\subsection{VLT/MUSE}
We make use of the PHANGS-MUSE survey (PI: E.~Schinnerer; Emsellem et al.\ in prep.). This survey employs the Multi-Unit Spectroscopic Explorer \citep[MUSE;][]{Bacon2014} optical integral field unit (IFU) mounted on the VLT UT4 to mosaic the star-forming disk of $19$ galaxies from the PHANGS sample. These galaxies correspond to a subset of the $74$ galaxies from the PHANGS-ALMA survey \citep[PI: E.~Schinnerer;][]{Leroy2021b}. However, we have excluded one galaxy from the PHANGS-MUSE sample (NGC~0628) because its MUSE mosaic was obtained using a different observing strategy, leading to differences in data quality. The target selection for the PHANGS-MUSE sample focused on galaxies from the PHANGS parent sample that had already available ALMA data, as part of the ALMA pilot project, or from the ALMA archival. 

The mosaics consist of $3$ to $15$ individual MUSE pointings. Each pointing provides a $1\arcmin \times 1\arcmin$ field of view sampled at $0\farcs2$ per pixel, with a typical spectral resolution of ${\sim}2.5$~\AA\ (${\sim}40$~km~s$^{-1}$) covering the wavelength range of $4800{-}9300$~\AA. The total on-source exposure time per pointing for galaxies in the PHANGS-MUSE Large Program is $43$~min.  Nine out of the $18$ galaxies were observed using wide-field adaptive optics (AO). These galaxies are marked with a black dot in the first column of Table~\ref{tab:sample}. The spatial resolution ranges from ${\sim}0\farcs5$ to ${\sim}1\farcs0$ for the targets with and without AO, respectively. Observations were reduced using a pipeline built on {\sc esorex} and developed by the PHANGS team\footnote{\url{https://github.com/emsellem/pymusepipe}} (Emsellem et al.\ in prep.). The total area surveyed by each mosaic ranges from $23$ to $692$~kpc$^{2}$. Once the data have been reduced, we have used the PHANGS data analysis pipeline (DAP, developed by Francesco Belfiore and Ismael Pessa) to derive various physical quantities. The DAP will be described in detail in Emsellem et al.\ (in prep.). It consists of a series of modules that perform single stellar population (SSP) fitting and emission line measurements to the full MUSE mosaic. Some of these outputs are described in Secs.~\ref{sec:SSP_fit} and~\ref{sec:SFR}.

\subsection{ALMA}
\label{sec:alma}

The $18$ galaxies have \mbox{CO(2--1)} data from the PHANGS-ALMA Large Program \citep[PI: E.~Schinnerer;][]{Leroy2021b}. We used the ALMA $12$m and $7$m arrays combined with the total power antennas to map CO emission at a spatial resolution of $1\farcs0 {-} 1\farcs5$. The CO data cubes have an rms noise of ${\sim}0.1$~K per $2.5$~km~s$^{-1}$ channel (corresponding to $1\,\sigma(\Sigma_\mathrm{mol.gas}) \approx 2$~M$_{\odot}$~pc$^{-2}$ per $5$~km~s$^{-1}$ interval). The inclusion of the Atacama Compact Array (ACA) $7$m and total power data means that these maps are sensitive to emission at all spatial scales. For our analysis we use the integrated intensity maps from the broad masking scheme, which are optimized for completeness and contains the entirety of the galaxy emission.
The strategy for observing, data reduction and product generation are described in \citet{Leroy2021a}.
As our fiducial $\alpha_\mathrm{CO}$ conversion factor we adopt the local gas-phase metallicity in solar units ($Z' \equiv Z / Z_\odot$) scaled prescription as described in \citet{Accurso2017} and \citet{Sun2020}, that is $\alpha_\mathrm{CO} = 4.35 Z'^{-1.6}$ M$_{\odot}$~pc$^{-2}$ (K~km~s$^{-1}$)$^{-1}$, adopting a ratio CO(2-1)-to-CO(1-0) = 0.65 \citep[][T.~Saito et al.\ in prep.]{Leroy2013, denBrok2021}. The radially-varying metallicity is estimated from the radial profile of gas-phase abundances in \hii\ regions, as explained in  \citet[][]{Kreckel2020}. Azimuthal variations in the metallicity of the interstellar medium have been previously reported \citep{Ho2017, Kreckel2020}, however, these variations are small ($0.04 - 0.05$ dex), implying variations of $\sim0.06$ dex in $\alpha_{\mathrm{CO}}$) and therefore do not impact our results. We test the robustness of our results against a constant $\alpha_\mathrm{CO} = 4.35$ M$_{\odot}$~pc$^{-2}$ (K~km~s$^{-1}$)$^{-1}$, as the canonical value for our Galaxy \citep{Bolatto2013} in Sec.~\ref{sec:conv_fact}. IC~5332 has no significant detection of \mbox{CO(2--1)} emission and therefore has been excluded from the rKS and rMGMS analysis.

\begin{table*}
\centering
\begin{center}
\renewcommand{\arraystretch}{1.2}
\begin{tabular}{lcccccrccc}
\hline
\hline
Target & RA & DEC & log$_{10}\mathrm{M}_{*}$ & log$_{10}\mathrm{M}_{\mathrm{H}_{2}}$ & log$_{10}{\rm SFR}$ & $\Delta$MS & Distance & Inclination & Mapped area \\
 & (degrees) & (degrees) & $(M_{\odot})$ & $(M_{\odot})$ & $(M_{\odot}\,yr^{-1})$ & (dex) & (Mpc) & (degrees) & (kpc$^{2})$ \\
 (1) & (2) & (3) & (4) & (5) & (6) & (7) & (8) & (9) & (10)\\
\hline 
NGC~1087 & $41.60492$ & $-0.498717$ & 9.9 & $9.2$ & $0.12$ & $0.33$ & 15.85$\pm$2.08 & 42.9 & 128 \\
NGC~1300$^{\bullet}$ & $49.920815$ & $-19.411114$ & 10.6 & $9.4$ & $0.07$ & $-0.18$ & 18.99$\pm$2.67 & 31.8 & 366 \\
NGC~1365 & $53.40152$ & $-36.140404$ & 11.0 & $10.3$ & $1.23$ & $0.72$ & 19.57$\pm$0.77 & 55.4 & 421 \\
NGC~1385$^{\bullet}$ & $54.369015$ & $-24.501162$ & 10.0 & $9.2$ & $0.32$ & $0.5$ & 17.22$\pm$2.42 & 44.0 & 100 \\
NGC~1433$^{\bullet}$ & $55.506195$ & $-47.221943$ & 10.9 & $9.3$ & $0.05$ & $-0.36$ & 18.63$\pm$1.76 & 28.6 & 441 \\
NGC~1512 & $60.975574$ & $-43.348724$ & 10.7 & $9.1$ & $0.11$ & $-0.21$ & 18.83$\pm$1.78 & 42.5 & 270 \\
NGC~1566$^{\bullet}$ & $65.00159$ & $-54.93801$ & 10.8 & $9.7$ & $0.66$ & $0.29$ & 17.69$\pm$1.91 & 29.5 & 212 \\
NGC~1672 & $71.42704$ & $-59.247257$ & 10.7 & $9.9$ & $0.88$ & $0.56$ & 19.4$\pm$2.72 & 42.6 & 255 \\
NGC~2835 & $139.47044$ & $-22.35468$ & 10.0 & $8.8$ & $0.09$ & $0.26$ & 12.22$\pm$0.9 & 41.3 & 88 \\
NGC~3351 & $160.99065$ & $11.70367$ & 10.4 & $9.1$ & $0.12$ & $0.05$ & 9.96$\pm$0.32 & 45.1 & 76 \\
NGC~3627 & $170.06252$ & $12.9915$ & 10.8 & $9.8$ & $0.58$ & $0.19$ & 11.32$\pm$0.47 & 57.3 & 87 \\
NGC~4254$^{\bullet}$ & $184.7068$ & $14.416412$ & 10.4 & $9.9$ & $0.49$ & $0.37$ & 13.1$\pm$1.87 & 34.4 & 174 \\
NGC~4303$^{\bullet}$ & $185.47888$ & $4.473744$ & 10.5 & $9.9$ & $0.73$ & $0.54$ & 16.99$\pm$2.78 & 23.5 & 220 \\
NGC~4321$^{\bullet}$ & $185.72887$ & $15.822304$ & 10.7 & $9.9$ & $0.55$ & $0.21$ & 15.21$\pm$0.49 & 38.5 & 196 \\
NGC~4535$^{\bullet}$ & $188.5846$ & $8.197973$ & 10.5 & $9.6$ & $0.33$ & $0.14$ & 15.77$\pm$0.36 & 44.7 & 126 \\
NGC~5068 & $199.72807$ & $-21.038744$ & 9.4 & $8.4$ & $-0.56$ & $0.02$ & 5.2$\pm$0.22 & 35.7 & 23 \\
NGC~7496$^{\bullet}$ & $347.44702$ & $-43.42785$ & 10.0 & $9.3$ & $0.35$ & $0.53$ & 18.72$\pm$2.63 & 35.9 & 89 \\
IC5332 & $353.61453$ & $-36.10108$ & 9.7 & $-$ & $-0.39$ & $0.01$ & 9.01$\pm$0.39 & 26.9 & 34 \\
\hline
\end{tabular}
\end{center}
\caption{Summary of the galactic parameters of our sample adopted through this work. $^{\bullet}$: Galaxies observed with MUSE-AO mode. Values in columns (4), (5) and (6) correspond to those presented in \citet{Leroy2021b}. Column (7) shows the vertical offset of the galaxy from the integrated main sequence of galaxies, as defined in \citet{Leroy2019}. Distance measurements are presented in \citet{Anand2021} and inclinations in \citet{Lang2020}. Uncertainties in columns (4), (5), (6) and (7) are on the order of $0.1$ dex. Column (10) shows the area mapped by MUSE.}
\label{tab:sample}
\end{table*}

\subsection{Environmental masks}
\label{sec:enviromental_mask}
We have used the environmental masks described in \citet{Querejeta2021} to morphologically classify the different environments of each galaxy and label them as disk, spiral arms, rings, bars and centers. This classification was done using photometric data mostly from the Spitzer Survey of Stellar structure in Galaxies \citep[S$^4$G;][]{Sheth2010}.
In brief, disks and centers are identified via 2D photometric decompositions of $3.6~\mu$m images \citep[see, e.g.,][]{Salo2015}. A central excess of light is labeled as center, independently of its surface brightness profile. The size and orientation of bars and rings are defined visually on the NIR images; for S$^4$G galaxies we follow \citet{Herrera-Endoqui2015}. Finally, spiral arms are only defined when they are clearly dominant features across the galaxy disk (excluding  flocculent spirals). First, a log-spiral function is fitted to bright regions along arms on the NIR images, and assigned a width determined empirically based on CO emission. For S$^4$G, we rely on the analytic log-spiral segments from \citet{Herrera-Endoqui2015}, and performed new fits for the remaining galaxies. 
These environmental masks allow us to examine the variations of the SFMS, rKS, and rMGMS relations across different galactic environments.

\subsection{Stellar mass surface density maps} 
\label{sec:SSP_fit}

The PHANGS-MUSE DAP (Emsellem et al. in prep) includes a stellar population fitting module, a technique where a linear combination of SSP templates of known ages, metallicities, and mass-to-light ratios is used to reproduce the observed spectrum. This permits us to infer stellar population properties from an integrated spectrum, such as mass- or light-weighted ages, metallicities, and total stellar masses, together with the underlying star formation history (which will be used in Sec.~\ref{sec:SFR_from_SFH}). Before doing the SSP fitting, we correct the full mosaic for Milky Way extinction assuming a \citet[][]{Cardelli89} extinction law and the $E(B{-}V)$ values obtained from the NASA/IPAC Infrared Science Archive\footnote{\url{https://irsa.ipac.caltech.edu/applications/DUST/}} \citep{Schlafly2011}.
In detail, our spectral fitting pipeline performs the following steps:
First, we used a Voronoi tessellation \citep{Capellari2003} to bin our MUSE data to a minimum signal-to-noise ratio (S/N) of ${\sim}35$, computed at the wavelength range of $5300{-}5500$~\AA. We chose this value in order to keep the relative uncertainty in our mass measurements below $15\%$, even for pixels dominated by a younger stellar population. To do this, we tried different S/N levels to bin a fixed region in our sample, and we bootstrapped our data to have an estimate of the uncertainties at each S/N level.

We use then the Penalized Pixel-Fitting (pPXF) code \citep{Capellari2004, Capellari2017} to fit the spectrum of each Voronoi bin. To fit our data, we used a grid of templates consisting of $13$ ages, ranging from $30$~Myr to  $13.5$~Gyr, logarithmically-spaced, and six metallicity bins $\mathrm{[Z/H]} = [-1.49,\ -0.96,\ -0.35,\ +0.06,\ +0.26,\ +0.4]$. We fit the wavelength range $4850{-}7000$~\AA, in order to avoid spectral regions strongly affected by sky residuals. We used templates from the eMILES \citep{Vazdekis2010, Vazdekis2012} database, assuming a \citet{Chabrier2003} IMF and BaSTI isochrone \citep{Pietrinferni2004} with a Galactic abundance pattern.  

The SSP fitting was done in two steps. First, we fitted our data assuming a \citet{Calzetti2000} extinction law to correct for internal extinction. We then corrected the observed spectrum using the measured extinction value before fitting it a second time, including a $12$ degree multiplicative polynomial in this iteration in the fit. This two-step fitting process accounts for offsets between individual MUSE pointings. The different MUSE pointings are not necessarily observed under identical weather conditions, and therefore, even with careful treatment, we find some systematic differences between the individual MUSE pointings related to the different sky continuum levels. We studied regions in our mosaic where different pointings overlap, and found variations on the order of ${\sim}3\%$ (between an identical region in two different pointings). After inducing similar perturbations on a subset of spectra, we found that even these small differences could potentially cause systematic differences in measured stellar-population parameters.
Therefore, in the first iteration of the SSP fitting, we measure a reddening value, and in the second iteration, we use a high-degree multiplicative polynomial to correct for those nonphysical features and homogenize the outcome of the different pointings.
Additionally, we have identified foreground stars as velocity outliers in the SSP fitting and we have masked those pixels out of the analysis carried out in this paper.

\subsection{Star formation rate measurements}
\label{sec:SFR}

As part of the PHANGS-MUSE DAP (Emsellem et al. in prep.), we fit Gaussian profiles to a number of emission lines for each pixel of the final combined MUSE mosaic of each galaxy in our sample. By integrating the flux of the fitted profile in each pixel, we were able to construct emission lines flux maps for every galaxy. In order to calculate our final SFR rate measurement we use the H$\alpha$, H$\beta$, \sii~and \oiii~emission line maps.
We de-reddend the H$\alpha$ fluxes, assuming that $\mathrm{H}\alpha_\mathrm{corr} / \mathrm{H}\beta_\mathrm{corr} = 2.86$, as appropriate for a case~B recombination, temperature $T = 10^{4}$~K, and density $n_\mathrm{e} = 100$~cm$^{2}$, following:

\begin{equation}
    \mathrm{H}\alpha_\mathrm{corr} = \mathrm{H}\alpha_\mathrm{obs}  
    \bigg(\frac{(\mathrm{H}\alpha / \mathrm{H}\beta)_\mathrm{obs}}{2.86}\bigg)^{\frac{k_{\alpha}}{k_{\beta} - k_{\alpha}}}~,
\end{equation}
where H$\alpha_\mathrm{corr}$ and H$\alpha_\mathrm{obs}$ correspond to the extinction-corrected and observed H$\alpha$ fluxes, respectively, and $k_{\alpha}$ and $k_{\beta}$ are the values of reddening in a given extinction curve at the wavelengths of H$\alpha$ and H$\beta$.  Opting for an \citet{ODonnell1994} extinction law, we use $k_{\alpha} = 2.52$, $k_{\beta} = 3.66$, and $R_\mathrm{V} = 3.1$.

Next, we determine whether the H$\alpha$ emission comes from gas ionized by recently born stars or by a different source, such as active galactic nuclei (AGN), or low-ionization nuclear emission-line regions (LINER). 
We performed a cut in the Baldwin–Phillips–Terlevich \citep[BPT;][]{BPT} diagram using the [\oiii]/H$\beta$ and [\sii]/H$\alpha$ line ratios, as described in \citet{Kewley2006}, to remove pixels that are dominated by AGN ionization from our sample. In the remaining pixels, we determined the fraction $C_\hii$ of the H$\alpha$ emission actually tracing local star formation, and the fraction deemed to correspond to the diffuse ionized gas (DIG), a warm ($10^{4}$~K), low density ($10^{-1}$~cm$^{-3}$) phase of the interstellar medium \citep{Haffner2009, Belfiore2015} produced primarily by photoionization of gas across the galactic disk by photons that escaped from \hii\ regions \citep[][F.~Belfiore et al.\ in prep]{Flores-Fajardo2011, Zhang2017}. To this end, we followed the approach described in \citet{Blanc2009} with the modifications introduced in \citet{Kaplan2016}. Essentially, we first use the [\sii]/H$\alpha$ ratio to estimate $C_\hii$ in each pixel, following:
\begin{equation}
\label{eq:CHII-def}
C_\hii = \frac{%
\displaystyle{%
\frac{[\sii]}{\mathrm{H}\alpha} - \left(\frac{[\sii]}{\mathrm{H}\alpha}\right)_\mathrm{DIG}
}}{%
\displaystyle{%
\left(\frac{[\sii]}{\mathrm{H}\alpha}\right)_\hii - \left(\frac{[\sii]}{\mathrm{H}\alpha}\right)_\mathrm{DIG}
}}~,
\end{equation}
where $\Big(\frac{\rm [\sii]}{\mathrm{H}\alpha}\Big)_\mathrm{DIG}$ and $\Big(\frac{\rm [\sii]}{\mathrm{H}\alpha}\Big)_\hii$ correspond to the typical [\sii]/H$\alpha$ ratio of DIG and \hii\ regions as measured in the faint (10th percentile) and bright (90th percentile) end of the H$\alpha$ distribution, respectively. Then, we perform a least squares fitting to find the best $\beta$ and $f_{0}$ parameters such that:
  \begin{equation}
\label{eq:CHII}
   C_\hii = 1.0 - \bigg(\frac{f_{0}}{\mathrm{H}\alpha}\bigg)^{\beta}~.
\end{equation}
Equation~\ref{eq:CHII} represents the fraction of H$\alpha$ emission tracing local star formation as a function of the H$\alpha$ flux in each pixel. For H$\alpha$ fluxes lower than $f_{0}$, the fraction is defined to be zero. This has been done for each galaxy separately.
Bright pixels, dominated by star formation ionization have $C_\hii \sim 1$, while fainter and DIG-contaminated pixels have lower values.
We include in Appendix~\ref{sec:ap_CHII} an example of the fitting of the parametrization defined in Eq.~\ref{eq:CHII} to our data.

Finally, we compute the H$\alpha$ emission tracing star formation (H$\alpha_{\hii}$) as $C_\hii\times$ H$\alpha$ in each pixel, while the fraction $1 - C_\hii$ is deemed to be DIG emission (H$\alpha_\mathrm{DIG}$). We then calculate the total H$\alpha_\mathrm{DIG}$ to H$\alpha_{\hii}$ ratio ($f_\mathrm{DIG}$) and rescale the H$\alpha_{\hii}$ flux of all pixels by ($1 + f_\mathrm{DIG}$). This correction is performed because photons that ionize the DIG, originally escaped from \hii\ regions. It represents, therefore, a spatial redistribution of the H$\alpha$ flux.
This approach permits us to estimate a star formation rate even in pixels contaminated by non-star-forming emission. 
A S/N cut of 4 for H$\alpha$ and 2 for H$\beta$ was then applied before computing the star formation rate surface density map using Eq.~\ref{eq:sfr}. However, most of the low S/N pixels have $C_\hii \approx 0$, and therefore, the S/N cut does not largely impact our results. Pixels below this S/N cut, pixels with $C_\hii \leq 0$ or pixels where $\mathrm{H}\alpha_\mathrm{obs} / \mathrm{H}\beta_\mathrm{obs} < 2.86$, are considered non-detections (see Sec.~\ref{sec:fit}). In Sec.~\ref{sec:fit} we discuss the importance of non-detections (i.e., pixels with non-measured SFR) in our analysis. However, our main findings do not change qualitatively when we set to zero the SFR of
 these pixels (i.e., pixels that are not dominated by star-forming ionization according to the BPT criterion).
Additionally, in Sec.~\ref{sec:aperture} we discuss the impact of removing all H$\alpha$ emission not associated with morphologically-defined \hii\ regions.

To calculate the corresponding star formation rate from the H$\alpha$ flux corrected for internal extinction and DIG contamination, we adopted the prescription described in \citet{Calzetti-book}:
\begin{equation}
    \label{eq:sfr}
    \frac{\mathrm{SFR}} {\mathrm{M}_{\odot}\,\mathrm{yr}^{-1}} = 5.5\times10^{-42}\frac{\mathrm{H}\alpha_\mathrm{corr}}{\mathrm{erg\, s}^{-1}}~.
\end{equation}
This equation is scaled to a Kroupa universal IMF \citep{Kroupa2001}, however, differences with the Chabrier IMF assumed for the SSP fitting are expected to be small \citep{Kennicutt2012}. With these steps we obtain SFR surface density maps for each galaxy in our sample. We acknowledge that Eq.~\ref{eq:sfr} assumes a fully-sampled IMF, and that the lowest SFR pixels (especially at high spatial resolution) may not form enough stars to fully sample the IMF. Hence, the measured SFR is more uncertain in this regime . However, due to our methodology to bin the data (see Sec.~\ref{sec:fit}), the higher uncertainty in the low $\Sigma_\mathrm{SFR}$ regime has little effect in our analysis.

\section{Methods}
\label{sec:methods}

\subsection{Sampling the data at larger spatial scales}
\label{sec:degrade}

In order to probe the relations under study at different spatial scales, we have resampled our native resolution MUSE and ALMA maps to pixel sizes of $100$~pc, $500$~pc and $1$~kpc. The degrading of the data to larger spatial scales is done by only resampling into larger pixel, rather than performing a convolution before resampling. The $100$~pc pixels are generally larger than the native spatial resolution of the maps. The resampling has been done for each one of the three relevant quantities: stellar mass surface density, star formation rate surface density, and molecular gas mass surface density.
For stellar mass maps, the resampling was directly performed in the MUSE native resolution stellar mass surface density map, produced by the PHANGS-MUSE DAP.
Calculating a resampled star formation rate surface density map required the resampling of each one of the line maps used for the BPT diagnostic and extinction correction. The parameters $f_{0}$ and $\beta$ used for the DIG correction are calculated only at the native MUSE resolution, and do not change with spatial scale. This assumes that the typical DIG flux surface density does not vary with spatial scale, and the measurement at the native resolution is better constrained due to the higher number of pixels.  Once the emission line maps were resampled, pixels dominated by AGN ionization were dropped from the analysis. The remaining H$\alpha$ emission was then corrected by internal extinction and for DIG contamination as explained in Sec.~\ref{sec:SFR}. The DIG contamination correction is done independently at each spatial scale. 
For the molecular gas mass surface density, we have proceeded similarly to the stellar mass surface density. Additionally we have imposed a S/N cut of $1.5$ for the molecular gas mass surface density map after the resampling, dropping the faintest and most uncertain pixels.
For each one of the resampled quantities, we have also resampled the corresponding variance map to perform the S/N cut and report the corresponding uncertainty. Finally, we have corrected the maps by inclination, using a multiplicative factor of $\cos(i)$, where $i$ corresponds to the inclination of each galaxy, listed in Table~\ref{tab:sample}. The adopted inclinations correspond to those reported in \cite{Lang2020}.

\begin{figure*}
    \centering
    \includegraphics[width = \textwidth]{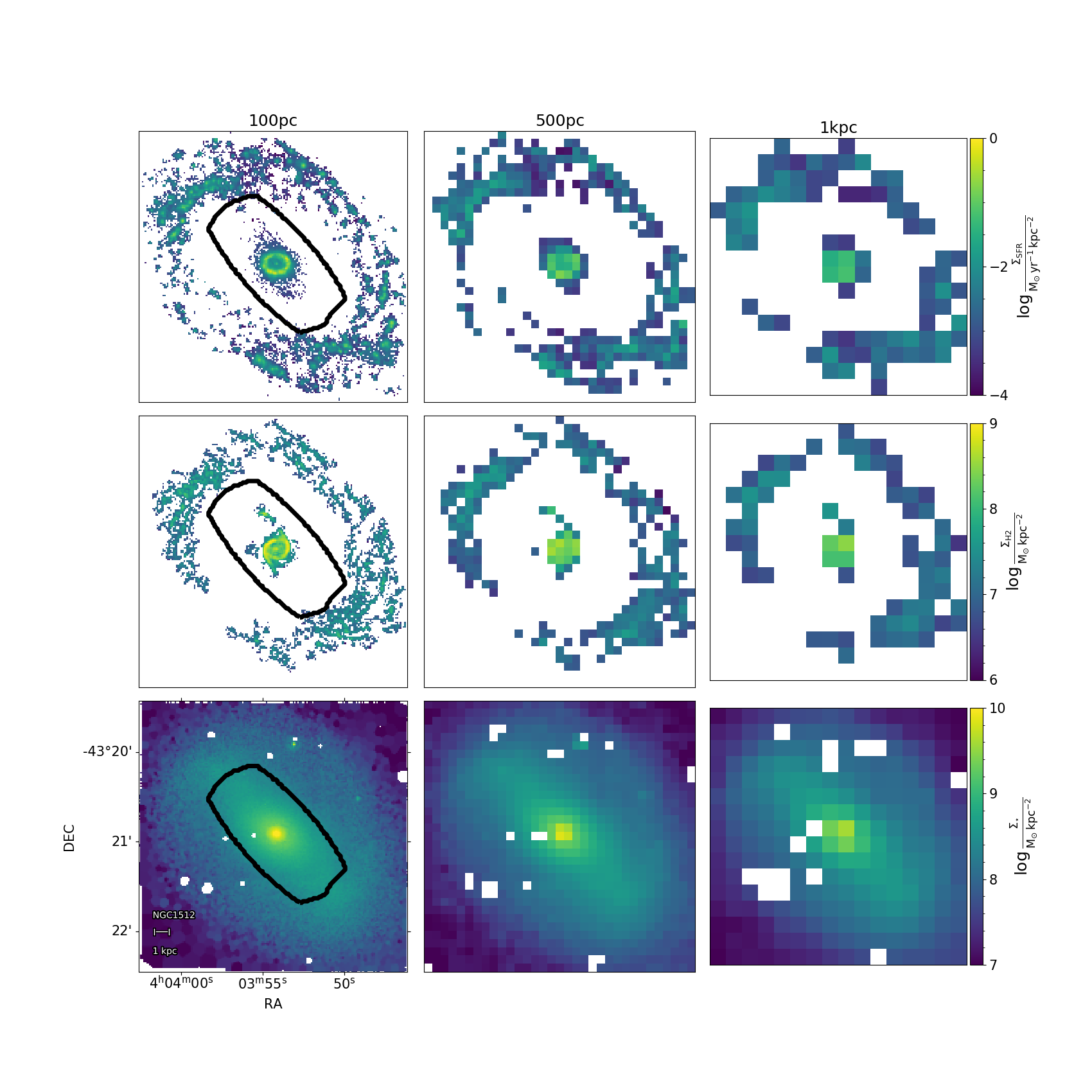}
    \caption{Example of SFR, molecular gas, and stellar mass surface density (top, middle, and bottom row) at spatial scales of $100$~pc, $500$~pc and $1$~kpc (left, middle, and right column) for one of the galaxies in our sample (NGC~1512). The black contour in each row encloses the pixels within the bar of the galaxy. Foreground stars have been masked in the stellar mass surface density maps (white pixels in bottom row).
    \label{fig:mapsNGC1512}}
\end{figure*}

Figure~\ref{fig:mapsNGC1512} shows the star formation rate surface density map (top), molecular gas mas surface density map (center), and stellar mass surface density map (bottom) at each of the spatial scales probed in this work ($100$~pc, $500$~pc, and $1$~kpc from left to right) for one example galaxy (NGC~1512).


\subsection{Fitting technique}
\label{sec:fit}

To fit each scaling relation, we binned the $x$-axis of our data in steps of $0.15$~dex, and calculated the mean within each bin. 
Fitting bins rather than single data points avoids giving statistically larger weight to the outer part of a galaxy, where more pixels are available. A minimum of 5 data points with nonzero signal per bin was imposed at the $100$~pc spatial scale for individual galaxies, in order to avoid sparsely sampled bins in the high or low end of the $x$-axis. For the full-sample relations, we imposed a minimum of 10 nonzero data points per bin at all spatial scales. We have also tested radial binning in the $x$-axis (i.e., calculating the means in bins defined as data points located at similar galactocentric radius), but this raised the scatter within each bin so we kept the $x$-axis binning, namely by stellar (gas) mass surface density.

The error in each bin has been calculated by bootstrapping: For a bin with $N$ data points, we repeatedly chose $100$ subsamples (allowing for repetitions), perturbed the data according to their uncertainties, and calculated their mean values. We adopt the standard deviation of the set of $100$ mean values as the uncertainty of the mean in that bin. We opted for this approach instead of standard error propagation because the uncertainties of our SFR measurements are too small, and this method offers a more conservative quantification of the scatter within each bin. Similarly, for each binning resultant from the $100$ iterations, we fit a power law and calculate a slope. To do this, we use a weighted least square fitting routine (WLS), where each bin is weighted by the inverse of its variance. This is justified since, by construction, the $x$-axis uncertainty of each bin is negligible, compared to its $y$-axis uncertainty.
The final slope and its error correspond to the mean and standard deviation of the slopes distribution, respectively. This quantification of the error accounts for sample variance and statistical uncertainty of each data point, but it does not reflect the uncertainties induced by systematic effects. Finally, the scatter reported throughout the paper for the fitted power laws corresponds to the median absolute deviation of each data point, considering detections only, with respect to the best-fitting power law.

However, at physical resolutions of ${\sim}100$~pc, we deal with the issue that within each bin, we observe (in the case of the rSFMS) a bimodal distribution of SFR. At a given stellar mass surface density, a fraction of the pixels probed have a nonzero value of SFR that correlates with its corresponding stellar mass surface density value (with a certain level of scatter), while the remaining pixels do not show any SFR within our detection limits. This is either because these pixels are intrinsically non-star-forming, or because their SFRs are lower than our detection threshold. 
The fraction of these ``non-detections'' (N/D) is higher in the lower stellar mass surface density regime and close to zero at the high-mass end. This bimodality reflects the fact that star formation is not uniformly distributed across galactic disks -- due to temporal stochasticity, or spatial organization of the star formation process due to galactic structure.

Thus, unlike studies done at ${\sim}$kpc resolutions, where star-forming regions smaller than the spatial resolution element will be averaged in a larger area, we need to properly account for the N/D fraction when investigating the scaling relations.
This requires designing a fitting method such that our measurements at high resolution will be consistent to those obtained at larger spatial scales.

For the analysis presented in this paper, we are interested in  (1) measuring how many stars are being formed per unit time on average at a given stellar mass surface density. This is different from asking (2) what is the typical SFR surface density at a given stellar mass surface density. The former requires us to include the non-detections as we are interested in averaging the SFR across the entire galactic disk, while the latter only tells us what are the most common SFR values to expect, and will depend on the probed spatial scale.

\begin{figure}
    \centering
    \includegraphics[width = \columnwidth]{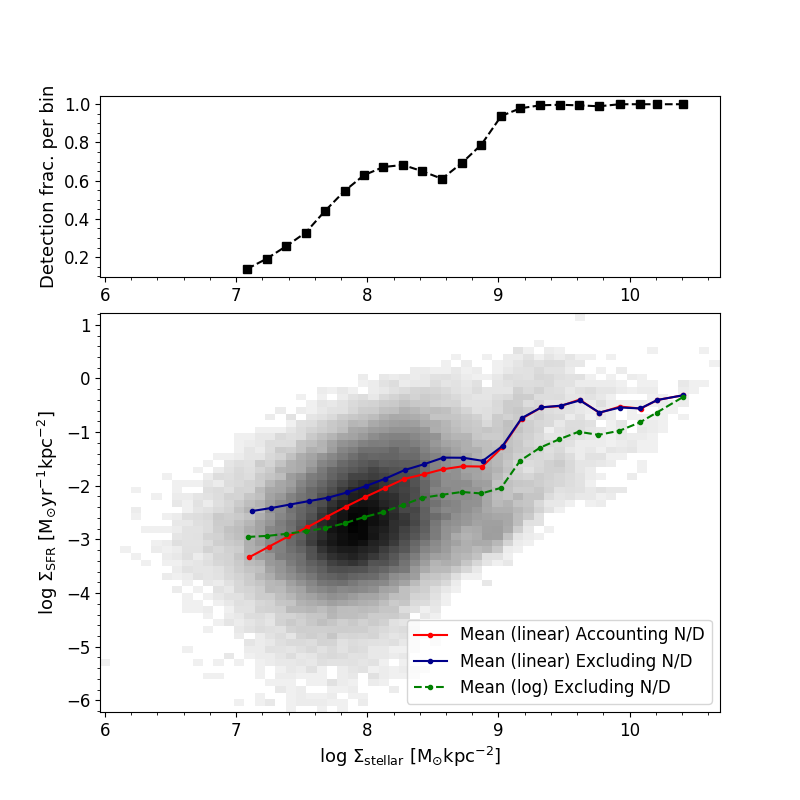}
    
    \caption{Bottom panel: 2D distribution of pixels in the overall rSFMS (i.e. including all pixels in our sample) and the three different binning schemes described in Sec.~\ref{sec:fit}. As explained in the main text, different binning schemes address different questions. The red line shows the fiducial binning scheme adopted in this paper. Top panel: Detection fraction of our SFR surface density tracer within each stellar mass surface density bin ranging from ${\sim}20\%$ in the low surface density regime to $100\%$ at the high surface density end. }
    
    \label{fig:bin-a}
\end{figure}

\begin{figure}
    \centering
    \includegraphics[width = \columnwidth]{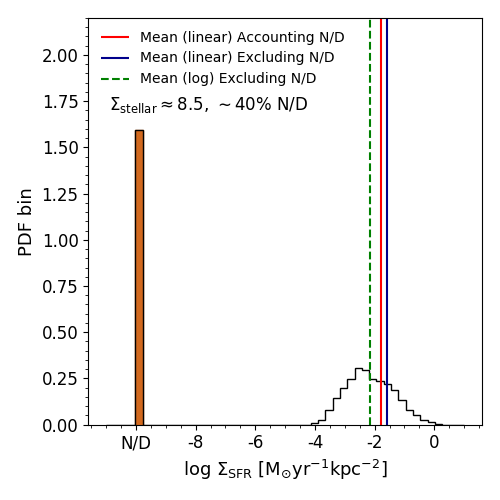}
    \caption{Distribution of the SFR surface density values within the mass bin of $\log \Sigma_\mathrm{stellar}\ [\mathrm{M}_\odot~\mathrm{pc}^{-2}]\approx 8.5$. Around $40\%$ of the pixels are not detected in $\Sigma_{\rm SFR}$ and we observe a bimodality in the distribution. These non-detections have been highlighted in brown and moved to an artificial value (N/D). The vertical lines correspond to the derived average value for the three binning schemes in Fig.~\ref{fig:bin-a}. The mean in log-space (green dashed line) matches the peak of the distribution of detections, while the mean in linear space (red and blue solid lines) correspond to the average SFR at this stellar mass surface density, including and excluding N/D, respectively.}
    
    \label{fig:bin-b}
\end{figure}

Measurements at lower spatial resolution are closer to addressing (1), as the physical interpretation of this is that in a given region of constant mass surface density, the mean SFR will be dominated by a few bright star-forming regions rather than by the much more numerous faint star-forming regions.
Figures~\ref{fig:bin-a} and~\ref{fig:bin-b} exemplify this difference for the rSFMS derived using all the available pixels in our data at a spatial resolution of ${\sim}100$~pc. Figure~\ref{fig:bin-a} shows three different binning schemes for the rSFMS in the bottom panel, and the detection fraction (defined as $1 - f_\mathrm{N/D}$) of each bin in the top panel, where $f_\mathrm{N/D}$ quantifies the fraction of N/D in each bin.

When averaging in linear space either including non-detections as zeros (red line) or excluding them from the average calculation (blue line) we address question (1), whereas averaging in log-space (green line) provides information on (2) and hence goes through the bulk of the 2D distribution in the $\log \Sigma_\mathrm{SFR}$ versus $\log \Sigma_\mathrm{stellar}$ plane.
Figure~\ref{fig:bin-b} shows the SFR distribution of pixels at an average $\log \Sigma_\mathrm{stellar}\ [\mathrm{M}_\odot\,\mathrm{pc}^{-2}] \approx 8.5$, and the average $\log \Sigma_\mathrm{SFR}$ for this stellar mass surface density bin computed with each one of the three binning schemes is shown by the vertical lines (using the same color scheme as for Fig.~\ref{fig:bin-a}). Non-detections in this bin are highlighted in brown. Here it becomes clear that while the mean in log-space (green line) matches the peak of the nonzero distribution, the means in linear space including or excluding non-detections (red and blue lines) are shifted toward higher values. Additionally the mean in linear space excluding non-detections (blue line) will always be greater or equal to the mean in linear space when non-detections are accounted for, as zeros in the calculation of the average (red line).

For our analysis we are interested in understanding if and how the star formation scaling relations (i.e., rSFMS, rKS and rMGMS) vary with measurement scale, and therefore we adopt the mean measured in linear space as our fiducial approach (i.e., red line in Fig.~\ref{fig:bin-a}). To probe the effect of excluding non-detections on the slope determination, we use the blue binning scheme (i.e., mean in linear space excluding N/D) in order to perform a fair comparison to the fiducial case. However, our results (in terms of impact of spatial scale in the measured slope) are qualitatively unchanged if we use the mean in log space (i.e., green line) in this case instead.

Finally, we highlight here that for our fiducial approach, we are interested in probing the scaling relations across the entire galactic disk. Hence, we consider all the pixels from a given galaxy in the computation of the scaling relations, including faint and more uncertain SFR measurements, as well as N/D. We note that a pixel is defined as N/D (in a given scaling relation) if it has a nonzero measurement in the quantity shown on the $x$-axis, and a value below our detection limits in the quantity shown on the $y$-axis, i.e. a non-detection in molecular gas will lead to the omission of this pixel in the rKS, while it will be treated as an N/D in the rMGMS.


\section{Results}
\label{sec:results}

In this section we present our measurements of the three relations under study. We recover all three scaling relations at a spatial scale of $100$~pc and we explore how these relations change when the data are degraded to lower spatial resolutions. We present each relation when considering all available pixels from the full galaxy sample, and derived for each individual galaxy to study galaxy-to-galaxy variations. As mentioned in Section~\ref{sec:alma}, IC~5332 has not been detected in \mbox{CO(2--1)} emission and therefore, the rKS and rMGMS have been measured for the remaining $17$ galaxies only.

\subsection{Scaling relations using the full sample}
\label{sec:global_results}

\begin{figure*}
    \centering
    \includegraphics[width = \textwidth]{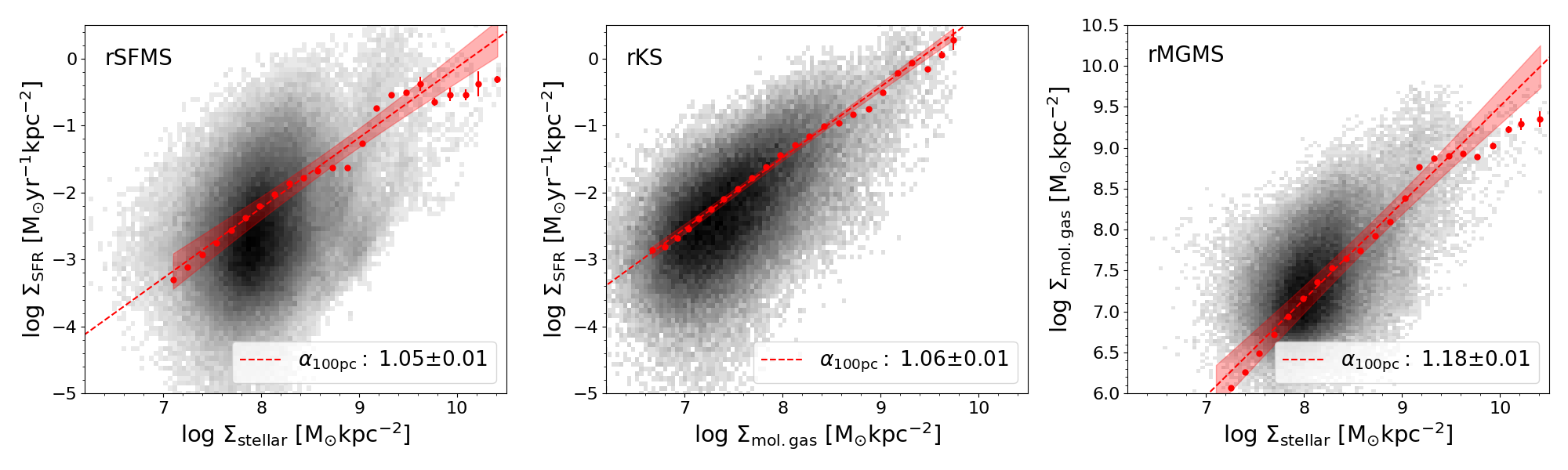}
    \caption{2D~distribution of the overall resolved star formation main sequence (left), resolved Kennicutt--Schmidt relation (center), and molecular gas main sequence (right) at $100$~pc spatial scale. The red points show the mean binned data and the red dashed line is the best-fitting power law. The colored region shows the $98\%$ confidence interval of the linear fit.
    \label{fig:global_all}}
\end{figure*}

Figure~\ref{fig:global_all} shows the 2D histograms of the overall rSFMS (left), rKS (center), and rMGMS (right), and their corresponding best-fitting power laws (red dashed line and red dots, respectively) at $100$~pc spatial scale.

The number of data points used is $313\,227$, $110\,084$ and $309\,569$ for the rSFMS, rKS, and rMGMS, respectively. This number is smaller for the rMGMS than for the rSFMS as IC~5332 is missing in the former. In the same order, the total detection fraction, defined as the fraction of pixels with a nonzero detection in the $y$-axis ($1-f_\mathrm{N/D}$), is $0.50$, $0.84$, and $0.35$. This means that about $16{-}65\%$ of the data points with a valid measurement of the property on the $x$-axis are not detected in the property on the $y$-axis. This reflects the high level of stochasticity dominating these relations at $100$~pc spatial scale.

Out of these three relations, we find that the rMGMS has the lowest scatter ($\sigma \approx 0.34$~dex), followed by the rKS and the rSFMS with scatter $\sigma \approx 0.41$~dex and $\sigma \approx 0.51$~dex, respectively (see bottom row in Table~\ref{tab:slopes_all_galaxies}). However, one must be careful when interpreting these numbers, since the scatter is computed using the nonzero pixels only. Due to our methodology, we include faint nonzero SFR pixels (see Sec.~\ref{sec:fit}) that enhance the scatter of the rSFMS and the rKS, especially at high resolution (see Sec.~\ref{sec:main_findings}). 


We have measured a slope of $1.05\pm0.01$ for the rSFMS, $1.06\pm0.01$ for the rKS, and $1.18\pm0.01$ for the rMGMS. In the same order, the intercepts that set the normalization of these relations are $-10.63$, $-9.96$, $-2.23$. These numbers are within the range of previously reported values, using lower resolution data. Differences in the slope possibly arise due to the use of different fitting methods \citep{Hsieh2017} or differences in the treatment of non-detection (see discussion in Sec.~\ref{sec:steepening}). Similarly, variations in the normalization of these relations are expected due to differences in the assumed CO-to-H$_{2}$ conversion factor (see Sec.~\ref{sec:conv_fact}), as well as in the binning methodology (see Sec.~\ref{sec:fit}). Figure~\ref{fig:bin-a} illustrates that the latter can lead to differences of up to $\sim0.5$ dex, with respect to the standard approach of average-in-log, depending on the fraction of N/D in a given bin. We summarize values from recent studies in Table~\ref{tab:previous_results}. Some studies reported more than one value, using different fitting methods, often orthogonal distance regression (ODR) and ordinary least squares (OLS). In that case we show here the OLS value, as we obtained similar numbers for both fitting techniques in our data, when non-detections are excluded from the analysis. 
The best-fitting slopes and scatter for the overall sample are provided in the bottom row of Table~\ref{tab:slopes_all_galaxies}.

\begin{table*}
\centering
\begin{tabular}{lrrrl}

\hline
\hline
Reference & $\alpha_{\mathrm{rSFMS}}$ & $\alpha_{\mathrm{rKS}}$ & $\alpha_{\mathrm{rMGMS}}$ & spatial scale \\
\hline
\citet{Sanchez2021} & $1.01\pm0.015$ & $0.95\pm0.21$ & $0.93\pm0.18$ & ${\sim}$kpc\\
\citet{Ellison2020} & $0.68\pm0.01$ & $0.86\pm0.01$ & $0.73\pm0.01$ & ${\sim}$kpc\\
\citet{Barrera-Ballesteros2021} & $0.92$ & $0.54$ & - &${\sim}2$ kpc\\
\citet{Morselli2020} & $0.74\pm0.25$ & $0.83\pm0.12$ & $0.94\pm0.29$ &${\sim}$500 pc\\
\citet{Lin2019} & $1.19\pm0.01$ & $1.05\pm0.01$ & $1.1\pm0.01$ &${\sim}$kpc\\
\citet{Dey2019}  & $-$ & $1.0\pm0.1$ & $-$&${\sim}$kpc\\
\citet{Medling2018}  & $0.72\pm0.04$ & $-$ & $-$&${\sim}$kpc\\
\citet{Abdurro2017} & $0.99$ & $-$ & $-$ &${\sim}$kpc\\
\citet{Hsieh2017}  & $0.715\pm0.001$ & $-$ & $-$&${\sim}$kpc\\
\citet{CanoDiaz2016}  & $0.72\pm0.04$ & $-$ & $-$&${\sim}$kpc\\
\citet{Leroy2013}  & $-$ & $1.0\pm0.15$& $-$ &${\sim}$kpc\\
\citet{Schruba2011}  & $-$ & $0.9\pm0.4$& $-$ &${\sim}0.2{-}2$ kpc\\
\citet{Blanc2009}  & $-$ & $0.82\pm0.05$& $-$ &${\sim}$750 pc\\
\citet{Bigiel2008}  & $-$ & $0.96\pm0.07$& $-$ &${\sim}$750 pc\\

\hline
\end{tabular}
\caption{Summary of some previously reported values for the slopes of the rSFMS, rKS law and rMGMS. The spatial scale at which each study was carried is indicated in the last column.}
\label{tab:previous_results}
\end{table*}

\subsection{Scaling relations in individual galaxies}
\label{sec:individual_gal}

\begin{figure}
    \centering
    \includegraphics[width = \columnwidth,height=12cm]{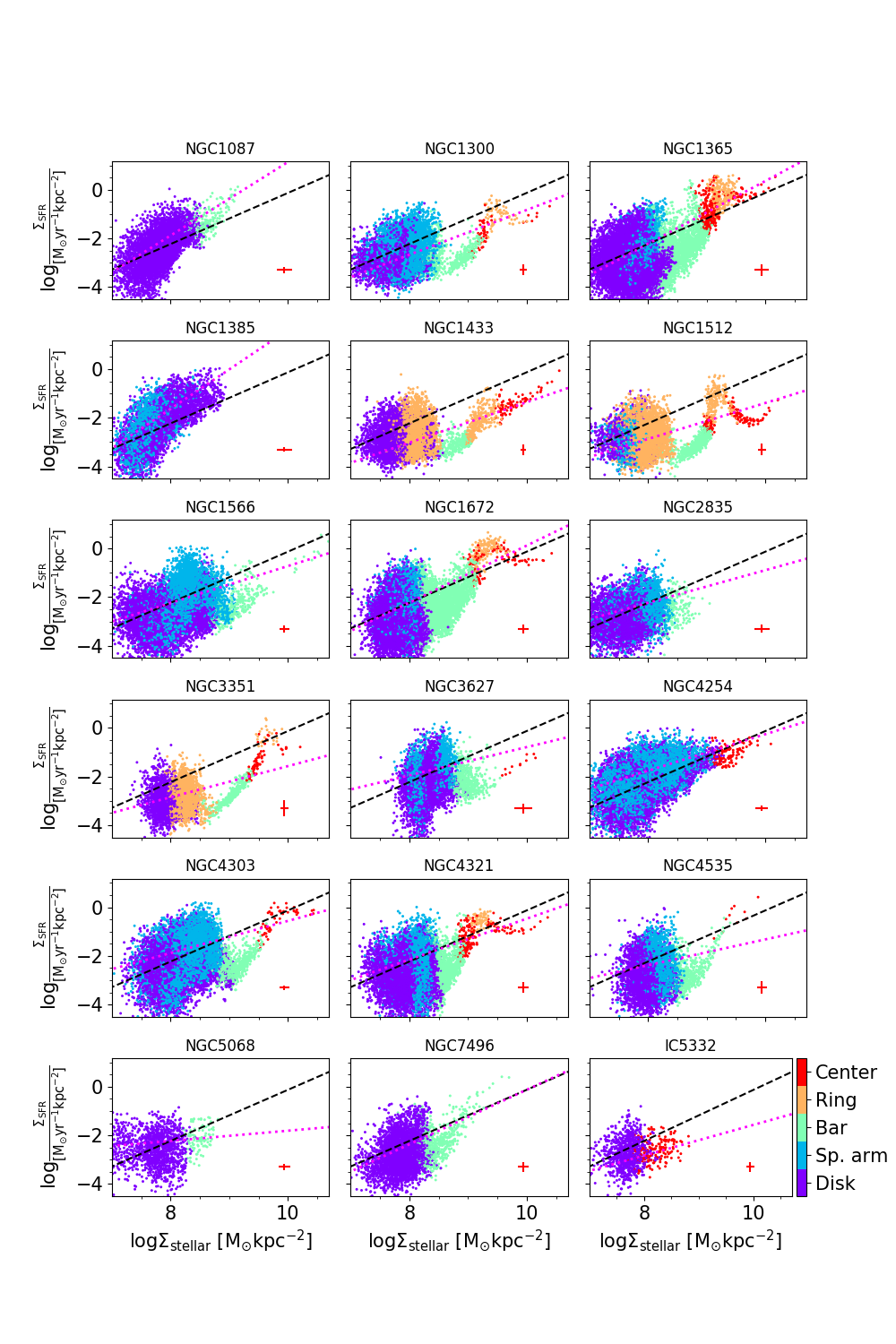}
    
    \caption{rSFMS for each galaxy at $100$~pc resolution. Different environment in the galaxies are colored according to the color scale next to the bottom right panel. The dashed black line represents the best-fitting power law to the overall measurement and the magenta line shows the best-fitting power law for each galaxy.}
    
    \label{fig:rSFMS_app100}
\end{figure}

\begin{figure}
    \centering
    \includegraphics[width=\columnwidth,height=12cm]{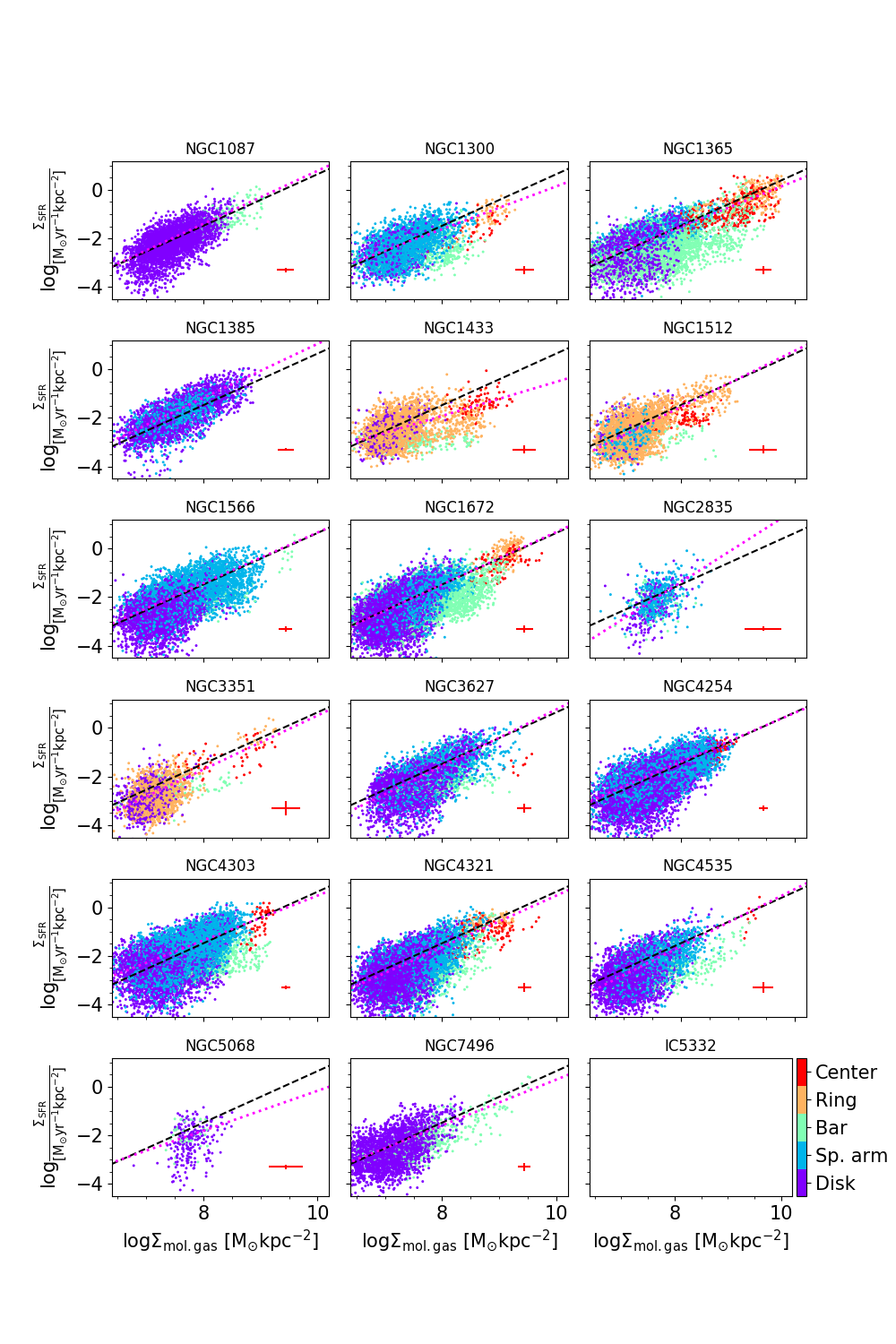}
    
    \caption{rKS for each galaxy at $100$~pc resolution. Different environments in the galaxies are colored according to the color scale next to the bottom right panel. The dashed black line represents the best-fitting power law to the overall measurement and the magenta line shows the best-fitting power law for each galaxy.}
    
    \label{fig:KS_app100}
\end{figure}

\begin{figure}
    \centering
    \includegraphics[width = \columnwidth,height=12cm]{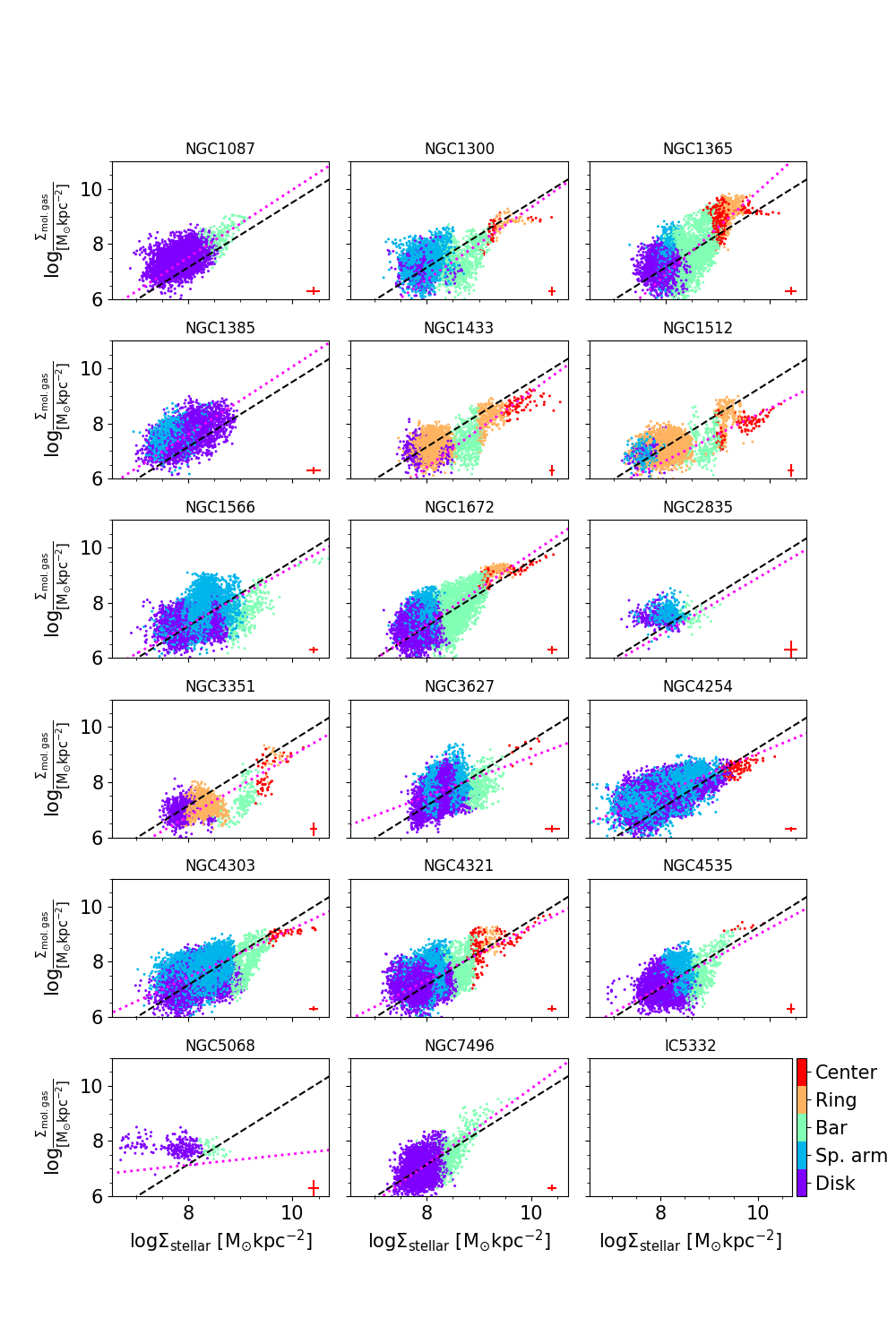}
    
    \caption{rMGMS for each galaxy at $100$~pc resolution. Different environments in the galaxies are colored according to the color scale next to the bottom right panel. The dashed black line represents the best-fitting power law to the overall measurement and the magenta line shows the best-fitting power law for each galaxy.}
    
    \label{fig:MolGas_app100}
\end{figure}

Here we explore galaxy-to-galaxy variations for the three scaling relations.
Figures~\ref{fig:rSFMS_app100}, \ref{fig:KS_app100}, and~\ref{fig:MolGas_app100} show the rSFMS, rKS, and rMGMS, respectively, for each individual galaxy in our sample.  Different colors represent different galactic environments, namely disk, spiral arms, bar, rings (inner and outer), and centers (see Sec.~\ref{sec:enviromental_mask}). The black dashed lines show the corresponding overall best-fitting power law from Fig.~\ref{fig:global_all} as reference and the magenta dashed line show the best-fitting power law for each individual galaxy. 
In Sec.~\ref{sec:global_results} we find that the rMGMS is the relation with the lowest scatter in our overall sample. Figure~\ref{fig:MolGas_app100} shows that the distribution of data points in this relation is much more compact (on a logarithmic scale) than in the rSFMS or the rKS. As mentioned earlier, this is due to the inclusion of very faint SFR pixels (see Sec.~\ref{sec:fit}). As explained in Sec.~\ref{sec:global_results}, the rMGMS is the relation that has the lowest total detection fraction in comparison. 

\begin{figure}
    \centering
    \includegraphics[width = 1.1\columnwidth]{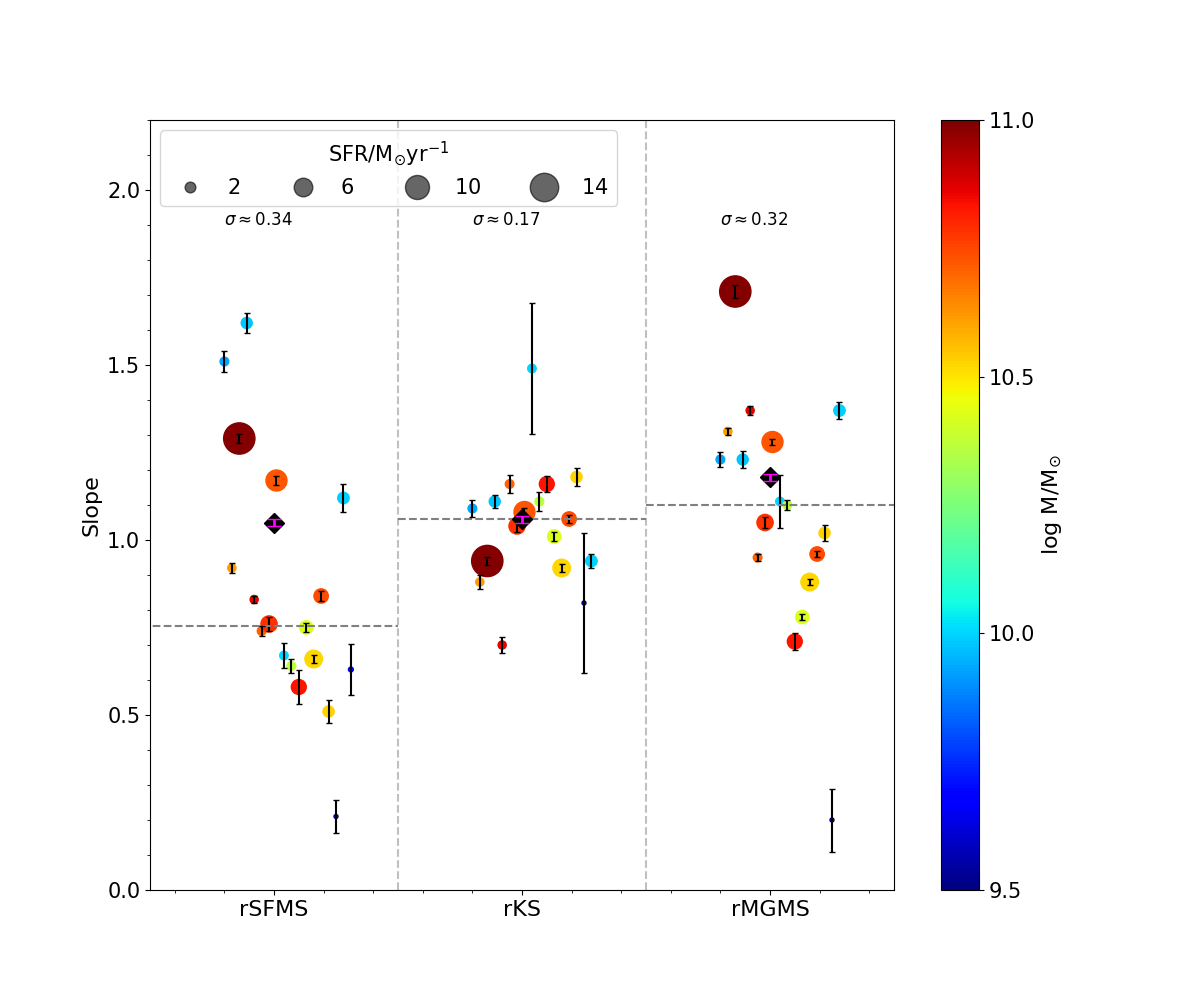}
    
    \caption{Slopes measured for the rSFMS, rKS and rMGMS for the individual galaxies in our sample. Each circle represents a galaxy. The size and color of each point scales with total SFR and stellar mass of the galaxy, respectively. The dispersion of the slope values for each relation is indicated on the top of the panel. The horizontal shift is arbitrary, with galaxies ordered from left to right by NGC number. The two galaxies with larger error bars in the KS law correspond to NGC~2835 and NGC~5068. This is due to the low number of data points available, as can be seen in Fig.~\ref{fig:KS_app100}. The black diamond with the magenta error bar indicates the global measurement for each relation. The horizontal gray dashed line show the median slope for each relation, in a galaxy basis.}
    
    \label{fig:slope_var}
\end{figure}

On the other hand, prominent differences between galaxies are seen for the rSFMS (Fig.~\ref{fig:rSFMS_app100}). For instance, in some galaxies such as NGC~1365 or NGC~1566, their disk and spiral arm pixels agree well with the overall rSFMS, while in others, like NGC~3351 or NGC~4321, disk, spiral arms, and outer ring pixels tend to exhibit a constant SFR surface density independent of their local stellar mass surface density (albeit probing only a limited range here).
Figure~\ref{fig:slope_var} highlights the diversity in the slopes measured for individual galaxies for the three relations probed.
The rSFMS relation shows the largest dispersion in slopes among galaxies ($\sigma\approx0.34$), followed by the rMGMS ($\sigma\approx0.32$), and finally the rKS ($\sigma\approx0.17$), where galaxies show slopes generally closer to the overall value (black diamond). The two galaxies with larger uncertainties in the rKS best-fitting power law slopes correspond to NGC~2835 and NGC~5068, two low-mass and low-SFR galaxies. Figure~\ref{fig:KS_app100} shows that these galaxies have only a small number of data points, hence their uncertain measurements.

Finally, we note that the shape of the rKS relation across different galaxies does not seem to change. While it has a larger scatter than the rMGMS at high resolution ($0.41$~dex versus $0.34$~dex, respectively), the relation is indeed more uniform across different environments. On the contrary, certain galactic environments do impact the shape of the rSFMS and the rMGMS. When present, inner structures (bar, inner ring, and center) often ``bend'' these scaling relations. In the case of the rSFMS, this ``bend'' might be related to an increase of SFR in an inner ring or a suppression of SFR in the bulge or bar, which is not reflected by a change in the stellar mass surface density but it is reflected in the molecular gas surface density. 

\begin{figure}
    \includegraphics[width = \columnwidth]{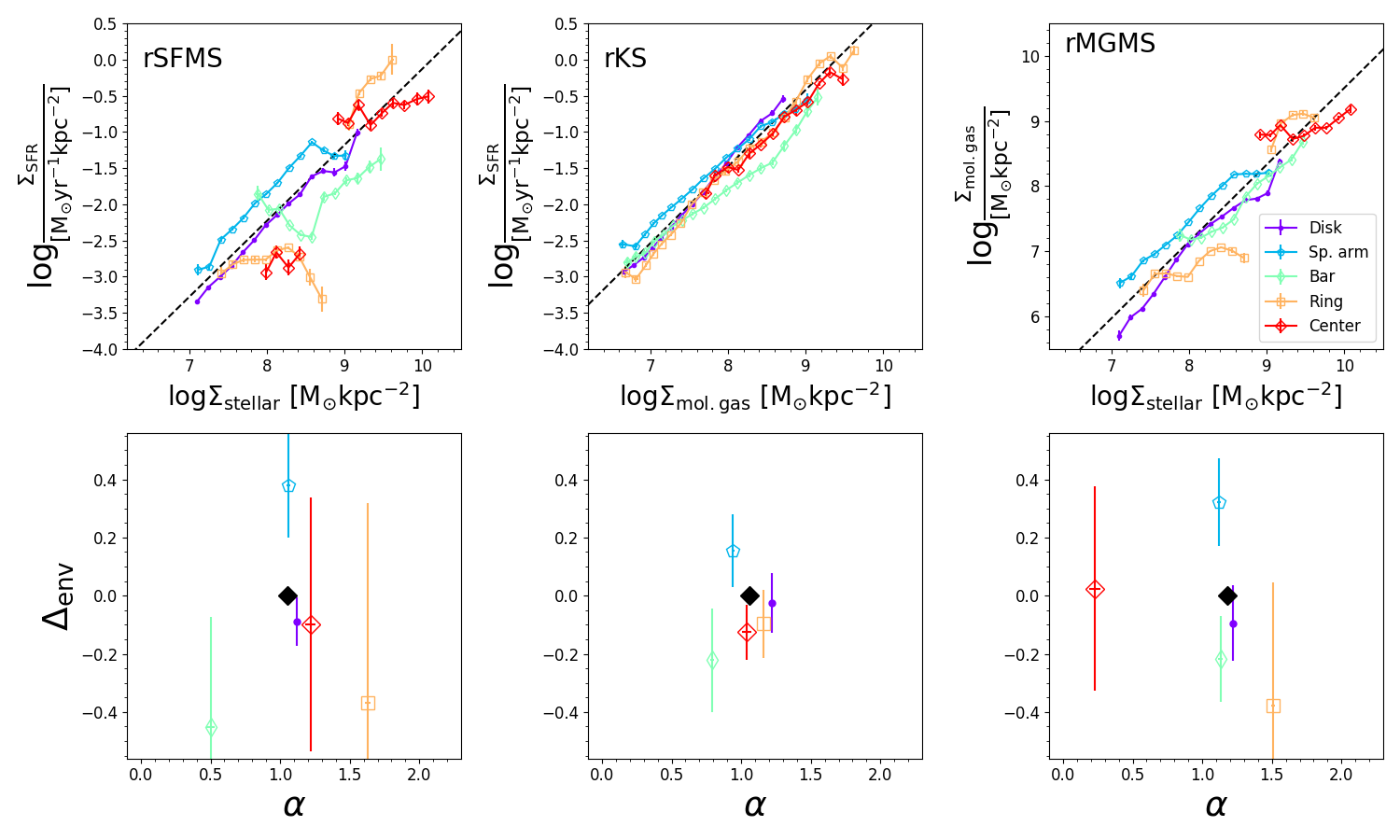}
    
    \caption{Variations across different galactic environments for the rSFMS (left), rKS (center) and rMGMS (right) relations. The top row shows the binned data for each galactic environment separately, following the same color scheme from Fig.~\ref{fig:MolGas_app100}, while the black dashed line shows the overall best-fitting power law to all data points together. The bottom row shows, on the $x$-axis the slope measured for each individual environment. The vertical offset between each binned environment and the overall best-fitting power law is reported on the $y$-axis. It has been calculated by fitting a power law with a fixed slope (that of the overall relation) to each binned environment, and computing the intercept difference respect to the overall best fitting power law. Its error-bars represent the standard deviation of the difference between the binned environment and the overall best fit. The black diamond show the slope of the overall best-fitting power law. (See Sec.~\ref{sec:individual_gal} for discussion).
}
    \label{fig:enviro}
\end{figure}

Figure~\ref{fig:enviro} highlights variations across different morphological environments. The top row shows binned scaling relations for each environment separately, considering all the pixels in our sample, compared to the overall relation we derived for all environments together. We use the offset ($\Delta_\mathrm{env}$) defined as the vertical offset between the binned scaling relation from each environment and the overall best-fitting power law, and the slope measured for each environment to quantify deviations from the overall relation. It can be seen that the slope and normalization of the rKS are similar across different galactic environments while the variations for the rSFMS and rMGMS relations are substantially larger (up to ${\sim}0.4$~dex). 

The disk environment (being the largest in area) is dominating the overall slope for rSFMS and rMGMS. The spiral arms share a similar slope, but are systematically offset above the overall relations by up to ${\sim}0.4$~dex. This is consistent with the findings reported in \citet{Sun2020}, where the authors found a systematic increase of $\Sigma_\mathrm{mol. gas}$ in spiral arms with respect to inter-arm regions analyzing $28$ galaxies from the PHANGS-ALMA survey with identified spiral arms. \citet{Querejeta2021} also report higher average $\Sigma_\mathrm{mol. gas}$ and $\Sigma_\mathrm{SFR}$ in spiral arms compared to inter-arm regions, using data from the PHANGS-ALMA survey and narrow-band H$\alpha$ imaging as SFR tracer. Outer rings (with low $\Sigma_\mathrm{stellar, mol. gas}$) appear in the rSFMS and rMGMS as flatter features laying below the overall relation, while inner rings (with high $\Sigma_\mathrm{stellar, mol. gas}$) show up above it. Finally, bars lie systematically below the overall relations, especially in the case of the rSFMS (see discussion in Sec.~\ref{sec:main_findings}). To our knowledge this is the first time these three scaling relations are probed separately across different morphological environments of galaxies, at a spatial scale comparable to individual star-forming regions.

\begin{figure}
    \centering
    \includegraphics[width = \columnwidth]{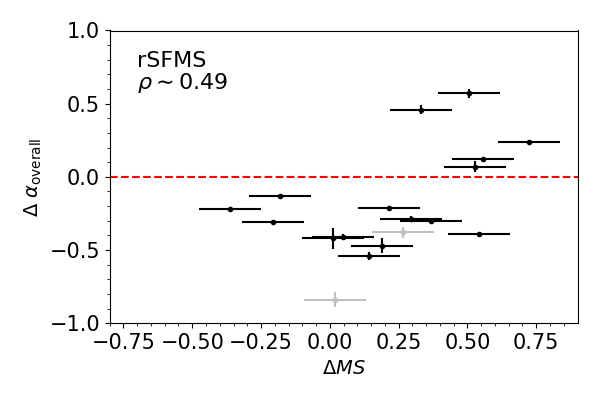}
    
    \caption{Differences in the slope measured for each individual galaxy with respect to the global measurements in Fig.~\ref{fig:global_all} for the rSFMS. Each dot represent a galaxy in our sample. The gray dots are NGC~2835 and NGC~5068, two low-mass galaxies. The PCC of the correlation is indicated in the top-left corner.}
    
    \label{fig:slope_var_vs_SFR}
\end{figure}

We investigated if any global galaxy parameter could be related to the diversity in the measured slopes. In particular, we explored any dependence of the slope on total stellar mass (M$_\star$), total SFR, specific SFR ($\mathrm{sSFR} = \mathrm{SFR} / \mathrm{M}_\star$), and vertical offset of the galaxy from the global star formation main sequence of galaxies ($\Delta$MS) as computed in \citet{Leroy2019}. Only correlations for $\Delta$MS are shown here, as the other parameters show even weaker or no correlations.
Figure~\ref{fig:slope_var_vs_SFR} shows the difference between the slope derived from the full sample and that of each individual galaxy ($\Delta\alpha_\mathrm{overall}$) as a function of $\Delta$MS of each galaxy. Specifically, if $\alpha_\mathrm{gal}$ corresponds to the slope measured for one individual galaxy, and $\alpha_\mathrm{overall}$ is the slope measured for the overall sample (see Table~\ref{tab:slopes_all_galaxies}), we define
\begin{equation}
\label{eq:global_def}
    \Delta\alpha_\mathrm{overall} = \alpha_\mathrm{gal} - \alpha_\mathrm{overall}~,
\end{equation}
Figure~\ref{fig:slope_var_vs_SFR} shows the slope variations with respect to $\Delta$MS for the rSFMS (An analog figure for the rKS and the rMGMS is included in the Appendix~\ref{sec:ap_deltaMS_rKSrMGMS} for completeness). NGC~2835 and NGC~5068 are marked as gray points. A weak correlation can be identified between the slope difference in the rSFMS and $\Delta$MS (Pearson correlation coefficient (PCC) of ${\sim}0.49$). A potential origin of this correlation is discussed in Sec.~\ref{sec:main_findings}. No correlation is found for the slope difference in the rKS or rMGMS with any of the tested global parameters. Table~\ref{tab:slopes_all_galaxies}
summarizes the slopes measured for each galaxy and each scaling relation probed.

\begin{table*}
\centering
\begin{tabular}{ccccccc}
\hline
\hline
Relation & $\alpha_{100\mathrm{pc}}$ & $\sigma_{100\mathrm{pc}}$ & $\alpha_{500\mathrm{pc}}$ & $\sigma_{500\mathrm{pc}}$ & $\alpha_{1\mathrm{kpc}}$ & $\sigma_{1\mathrm{kpc}}$ \\
\hline
rSFMS & $1.05\pm0.01$ & $0.51$ & $1.10\pm0.02$ & $0.49$ & $1.04\pm0.04$ & $0.44$ \\
rKS & $1.06\pm0.01$ & $0.41$ & $1.06\pm0.02$ & $0.33$ & $1.03\pm0.02$ & $0.27$ \\
rMGMS & $1.18\pm0.01$ & $0.34$ & $1.26\pm0.02$ & $0.32$ & $1.20\pm0.04$ & $0.29$ \\
\hline
\end{tabular}
\caption{Slope ($\alpha$) and scatter ($\sigma$) using all the available pixels in our sample, for each one of the scaling relations probed,  at spatial scales of $100$~pc, $500$~pc and $1$~kpc.}
\label{tab:global_slopes_allres}
\end{table*}
\begin{table*}
\centering
\begin{tabular}{ccccccc}
\hline
\hline

Object & $\alpha_{\mathrm{rSFMS}}$ & $\sigma_{\mathrm{rSFMS}}$ & $\alpha_{\mathrm{rKS}}$ & $\sigma_{\mathrm{rKS}}$ & $\alpha_{\mathrm{rMGMS}}$ & $\sigma_{\mathrm{rMGMS}}$ \\
\hline
NGC1087 & $1.51\pm0.03$ & $0.40$ & $1.09\pm0.02$ & $0.32$ & $1.23\pm0.02$ & $0.28$ \\
NGC1300 & $0.92\pm0.01$ & $0.40$ & $0.88\pm0.02$ & $0.38$ & $1.31\pm0.01$ & $0.33$ \\
NGC1365 & $1.29\pm0.01$ & $0.52$ & $0.94\pm0.01$ & $0.47$ & $1.71\pm0.02$ & $0.43$ \\
NGC1385 & $1.62\pm0.03$ & $0.44$ & $1.11\pm0.02$ & $0.32$ & $1.23\pm0.02$ & $0.30$ \\
NGC1433 & $0.83\pm0.01$ & $0.38$ & $0.70\pm0.02$ & $0.38$ & $1.37\pm0.01$ & $0.28$ \\
NGC1512 & $0.74\pm0.01$ & $0.42$ & $1.16\pm0.02$ & $0.37$ & $0.95\pm0.01$ & $0.25$ \\
NGC1566 & $0.76\pm0.02$ & $0.48$ & $1.04\pm0.02$ & $0.40$ & $1.05\pm0.02$ & $0.33$ \\
NGC1672 & $1.17\pm0.01$ & $0.49$ & $1.08\pm0.01$ & $0.42$ & $1.28\pm0.01$ & $0.32$ \\
NGC2835 & $0.67\pm0.04$ & $0.42$ & $1.49\pm0.19$ & $0.39$ & $1.11\pm0.08$ & $0.25$ \\
NGC3351 & $0.64\pm0.02$ & $0.38$ & $1.11\pm0.03$ & $0.36$ & $1.10\pm0.01$ & $0.22$ \\
NGC3627 & $0.58\pm0.05$ & $0.50$ & $1.16\pm0.02$ & $0.42$ & $0.71\pm0.02$ & $0.31$ \\
NGC4254 & $0.75\pm0.01$ & $0.47$ & $1.01\pm0.01$ & $0.37$ & $0.78\pm0.01$ & $0.29$ \\
NGC4303 & $0.66\pm0.01$ & $0.51$ & $0.92\pm0.01$ & $0.44$ & $0.88\pm0.01$ & $0.31$ \\
NGC4321 & $0.84\pm0.01$ & $0.50$ & $1.06\pm0.01$ & $0.40$ & $0.96\pm0.01$ & $0.28$ \\
NGC4535 & $0.51\pm0.03$ & $0.49$ & $1.18\pm0.03$ & $0.43$ & $1.02\pm0.02$ & $0.28$ \\
NGC5068 & $0.21\pm0.05$ & $0.45$ & $0.82\pm0.20$ & $0.47$ & $0.20\pm0.09$ & $0.19$ \\
NGC7496 & $1.12\pm0.04$ & $0.48$ & $0.94\pm0.02$ & $0.40$ & $1.37\pm0.02$ & $0.29$ \\
IC5332 & $0.63\pm0.07$ & $0.38$ & - & - & - & - \\
\hline
Overall & $1.05\pm0.01$ & $0.51$ & $1.06\pm0.01$ & $0.41$ & $1.18\pm0.01$ & $0.34$ \\
\hline
\end{tabular}
\caption{Slope ($\alpha$) and scatter ($\sigma$) for each galaxy in our sample, for each one of the three relations probed, at $100$~pc spatial scale. The overall measurement considering all the valid pixels is included in the last row. }
\label{tab:slopes_all_galaxies}
\end{table*}

\subsection{Variation of slope and scatter as a function of spatial scale}
\label{sec:spatial_scale_fiducial}
\begin{figure*}
    \centering
    \includegraphics[width = \textwidth]{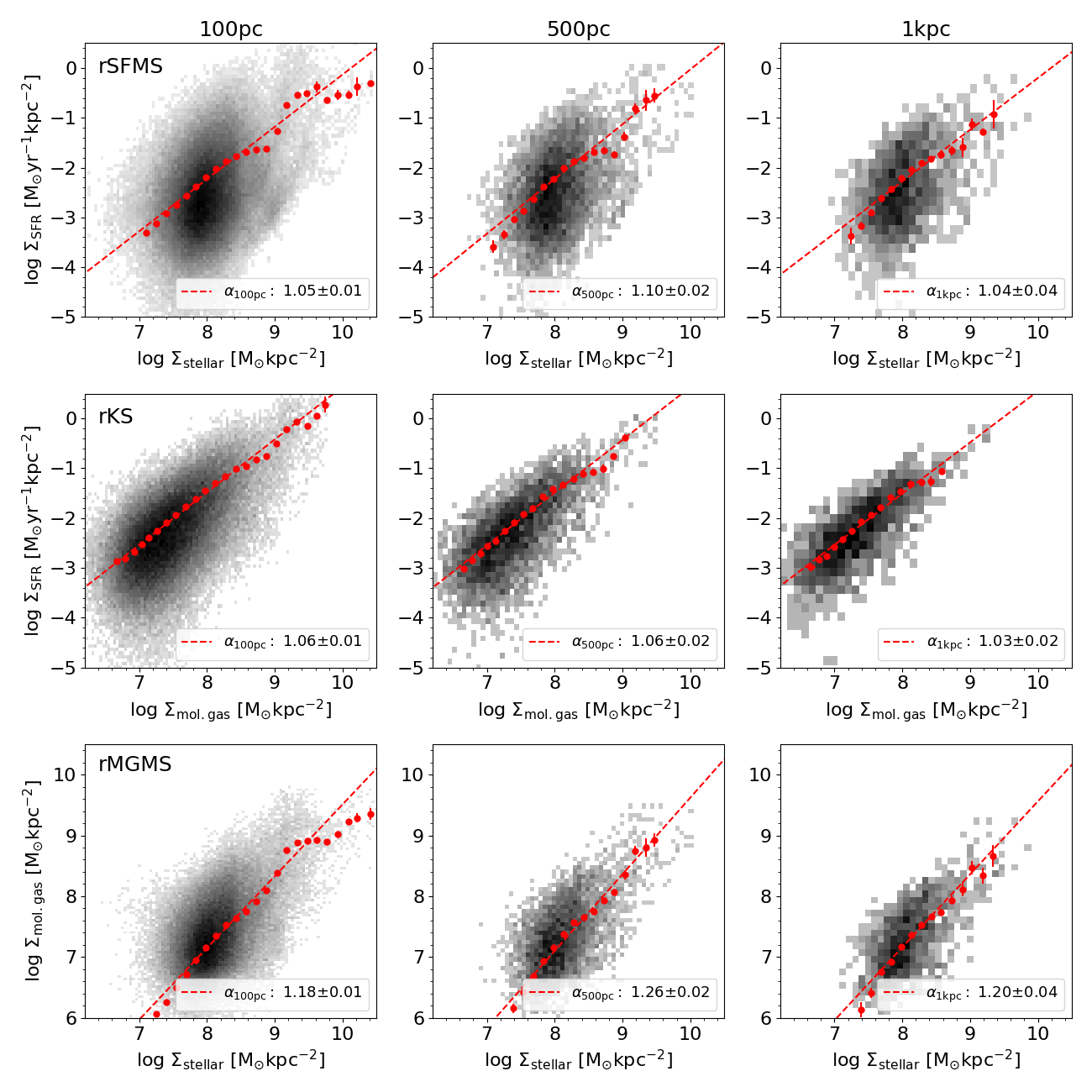}
    \caption{2D~histograms of the overall rSFMS (top row), rKS (center row), and rMGMS (bottom row) using all available pixels from our sample and probed at spatial scales of $100$~pc (left column), $500$~pc (middle column), and $1$~kpc (right column). The $x$-axis binning and the best-fitting power law are indicated with red dots and a red dashed lines, respectively. The measured slope and its error are stated in each panel.}
    \label{fig:variation_scale}
\end{figure*}

In this section, we explore how varying the spatial scale of the data affects the three relations. We resample our data to spatial scales of $500$~pc, and $1$~kpc as explained in Sec.~\ref{sec:degrade}. Figure~\ref{fig:variation_scale} shows the three relations probed at spatial scales of $100$~pc, $500$~pc, and $1$~kpc. 
The measured slopes and scatter are reported in Table~\ref{tab:global_slopes_allres}.
At all spatial scales, the slopes  show no evidence of systematic dependence with spatial resolution. Uncertainties in the slope measurements are larger at $1$~kpc due to the smaller number of data points, while the scatter is systematically lower at larger spatial scales, consistent with the findings reported in \citet{Bigiel2008, Schruba2010, Leroy2013, Kreckel2018}. This results from averaging small scales variations of regions at different stages of their evolutionary cycle \citep[e.g.,][]{Kruijssen2014}. In fact, at $1$~kpc resolution the rKS shows the lowest level of scatter.

Finally, the number of pixels used at $1$~kpc resolution drops to $2860$, $1510$, and $2820$, and the fraction of pixels with detected signal increases to $0.62$, $0.90$, and $0.52$ for the rSFMS, rKS, and rMGMS, respectively. This represents an increase of $24\%$, $7\%$, and $49\%$ with respect to the $100$~pc spatial scale.

\section{Discussion}
\label{sec:disc}

\subsection{Main findings of our analysis}
\label{sec:main_findings}

In the previous section, we presented our results on the scaling relations for both, our overall sample and individual galaxies. Our main findings are summarized in the following. 

\subsubsection{Variations with spatial scale between and within galaxies}

We find in our overall sample that the rMGMS is the relation with the smallest scatter (at a spatial scale of $100$~pc), followed by the rKS, and the rSFMS (see Table~\ref{tab:global_slopes_allres}). However, the rKS shows the most consistency across different galaxies and environments (i.e., similar values of slope and normalization). Its larger scatter is due to the inclusion of low SFR pixels that deviate more from the overall relation. The consistency of the rKS across different galaxies and environments reflects the direct connection between molecular gas and SFR, since stars form out of molecular gas. Hence, SFR ``follows'' the molecular gas distribution within short timescales \citep[H$\alpha$ emission traces star formation in the last ${\sim}10$~Myr;][]{Calzetti-book, Leroy2012, CatalanTorrecilla2015, Haydon2020} and this produces some correlated features in the shape of the rSFMS and the rMGMS for individual galaxies. NGC~1300, NGC~1512, and NGC~3351 are clear examples of these features. These similar features suggest that differences in the rSFMS across different environments are partially driven by the availability of molecular gas to fuel the formation of stars. However, bars, for example, generally appear below the overall best-fitting power law in the rKS as well, which suggests that variations in star formation efficiency may also play a role in driving the scatter.
This consistency (across different galaxies and environments) suggest that physically, it is the more fundamental relation, compared to the rSFMS and the rMGMS, in agreement with what has been found by \citet{Lin2019} and \citet{Ellison2020}. 

In contrast, from the perspective of individual galaxies, we see that the galactic environment has a stronger impact on the rSFMS and the rMGMS than on the rKS.  In particular, in the rSFMS, bars display a small range of SFR values, all systematically below the overall sequence. This apparent suppression of SFR in bars is consistent with the ``star formation desert'' \citep[][]{James2018} in barred galaxies.
Furthermore, variations in the rSFMS between different galaxies as those presented here have been previously reported \citep{Hall2018, Liu2018, Vulcani2019, Ellison2020}. Thus, a single power law across the full stellar mass surface density range does not provide a good representation of the $\Sigma_\mathrm{stellar}$ versus $\Sigma_\mathrm{SFR}$ plane; instead, different galactic environments may define different relations.
However, here we aim at measuring these scaling relations to first order, i.e., a ``simple'' power law, and finding a more realistic description of the data goes beyond the scope of this paper.

We also investigated if any global galaxy parameter relates to the galaxy-to-galaxy variations in the slope of the scaling relations studied, and we found a potential correlation between the slope of the rSFMS of a galaxy and its $\Delta$MS parameter. A positive correlation with a globally enhanced SFR could be explained as the increase of SFR happening preferably in the inner and more dense regions of the galaxy. This would lead to a steeper rSFMS in SFR-enhanced galaxies and it is consistent with the findings reported by \citet{Ellison2018}, where a global enhancement or suppression of SFR was found to impact the inner region of a galaxy stronger than its outskirts, studying a sample of galaxies from the MaNGA survey \citep{Bundy2015}. 
However, we stress here that even though the rSFMS shows hints of a correlation, the scatter and the limited number of data points make it hard to draw a robust conclusion.

Regarding spatial scale, we do not find evidence for a systematic dependence of the slope of these relations on the spatial resolution of the data, and the measurements are consistent within $2\sigma$ of their respective uncertainties. On the other hand, the scatter decreases at coarser spatial resolution as small scale variations are averaged out. The rKS is the relation that shows the least scatter at $1$~kpc spatial scale.

\subsubsection{Origin of observed scatter}
\label{sec:origin_scatter}
Numerous authors have investigated the physical origins of the scatter in these resolved scaling relations and determined how this scatter evolves with spatial scale. According to \citet{Schruba2010} and \citet{Kruijssen2014}, the scatter in the CO-to-H$\alpha$ ratio at small spatial scales is dominated by the fact that a given aperture can be either dominated by a peak in the CO emission (i.e. early in the star-forming cycle) or a peak in the H$\alpha$ emission (i.e. late in the star-forming cycle). At larger spatial scales, these peaks are averaged and this sampling effect diminishes. 

Along the same line, \citet{Semenov2017} presented a simple model to conciliate the long molecular gas depletion times measured at galactic scales \citep[${\sim}1{-}3$~Gyr;][]{Kennicutt1998, Bigiel2008, Leroy2008, Leroy2013} with the apparently shorter depletion times measured on $\sim100$~pc scales \citep[${\sim}40{-}500$~Myr;][]{Evans2009, Heiderman2010, Gutermuth2011, Evans2014, Schruba2017}. In the proposed scenario, the difference in these time scales originates from the fact that not all the gas is going through the star formation process at the same time. It is suggested that the fraction of molecular gas, that is actively forming stars, is regulated by local (stellar feedback, turbulence, gravitational instabilities) and global (large-scale turbulence, differential rotation) mechanisms that continuously turn on and off the formation of stars. Consequently, several of these star formation cycles must occur in order to process all the molecular gas in a given volume. Thus, at high spatial resolution, a larger scatter in the distribution of depletion times (and thus in the rKS relation) results from the decoupling of the actively star-forming clouds from the quiescent ones.

The local mechanism is discussed more generally in \citet{Kruijssen2014} and \citet{Kruijssen2018}, where the authors define a critical spatial scale on which the rKS breaks down from its integrated version, due to an incomplete sampling of the star-forming cycle. This spatial scale is defined as a function of the typical separation between independent star-forming regions, and the duration of the shortest phase of the star-forming process. This critical spatial scale equals the smallest region in which the star formation process is statistically well sampled. In this regard, having found the scatters at $100$ pc $\sigma_\mathrm{rSFMS} > \sigma_\mathrm{rKS} > \sigma_\mathrm{rMGMS}$ is consistent with the expectation from this evolutionary scenario, provided that $\tau_\mathrm{H\alpha} < \tau_\mathrm{CO} < \tau_\mathrm{stars}$, where $\tau$ corresponds to the duration each tracer is visible across the star formation cycle. Due to the fact that the period of time in which young stars ionize their surrounding interstellar medium is shorter than the lifetime of molecular clouds, i.e., $\tau_\mathrm{H\alpha} < \tau_\mathrm{CO}$ \citep{kruijssen2019,Chevance2020b}, the impact of time evolution on the rMGMS will be smaller than on the rSFMS or the rKS. Consequently, the minimum spatial scale needed to properly sample the rMGMS is smaller than that needed for the rSFMS or the rKS. At spatial scales of ${\gtrsim}1$~kpc, we are no longer in this ``undersampling'' regime, and the rKS shows the least scatter.

However, statistically incomplete sampling is not the only source of scatter for these resolved scaling relations. The slope of the rKS is also sensitive to the conditions present in the interstellar medium, such as metallicity and ultraviolet radiation field \citep{Feldmann2011}. Further \citet{Leroy2013} reported that 
variations in the measured depletion time ($\tau_\mathrm{dep} = \Sigma_\mathrm{mol. gas} / \Sigma_\mathrm{SFR}$) across different environments at ${\sim}$kpc spatial scales are consistent with variations of the CO-to-H$_{2}$ conversion factor. Similarly for the rSFMS, \citet{Ellison2020b} found that at ${\sim}$kpc spatial scales, the scatter in this relation is likely driven by local variations in  star formation efficiency. Here, we find large variations in these resolved scaling relation for different galaxies and galactic environments, in terms of both slope and normalization, which are particularly strong in the case of the rSFMS and rMGMS. These differences are key in setting their scatter, particularly at $1$~kpc spatial scale, where we are no longer in a stochastic regime, for the reasons described earlier.

\subsection{Role of non-detections in our analysis}
\label{sec:steepening}

In this section, we discuss the impact of excluding non-detections from the analysis. At high spatial resolution the non-detection fraction is quite large and their treatment can drastically alter the results. 

\subsubsection{Insights from simulations}

In \citet{Calzetti2012}, the authors studied variations in the slope of the observed rKS with different spatial resolution using simulated galaxies. They found that variations with spatial scale depend mainly on the slope of the true underlying rKS. In particular, for a slope of $1$, they found a nearly constant slope in the observed relation (variation of ${\sim}0.05$) from $200$~pc to $1$~kpc resolution. However, due to the assumed underlying close H$_{2}$--SFR relation, both quantities share the same spatial distribution and hence have basically identical detection fractions which is not necessarily the case for our data.

Similarly, in \citet{Hani2020} the authors used a sample of Milky Way-like galaxies from \mbox{FIRE-2} simulations to study the rSFMS measured at spatial scales of $100$~pc, $500$~pc, and $1$~kpc. They reported a systematic steepening of this relation at larger spatial scales. It is important to note that only pixels with nonzero SFR values were considered in their analysis. They interpreted that this effect was primarily caused by differences in the detection fraction of the SFR across the galactic disk. A lower detection fraction in outer regions, where the stellar mass surface density is lower, would cause a stronger dilution of the SFR in the low stellar mass surface density regime than in the inner and denser regions when spatially averaging the data. As a consequence, the rSFMS has been measured to be flatter at $100$~pc and to steepen at larger spatial scales, where the sparser intrinsic SFR distribution in the low mass surface density regime is leading to lower SFR surface density values. 

This steepening of the rSFMS is mainly a consequence of the methodology used (i.e., excluding N/D in the fitting process). In the following, we investigate this effect in our data, and measure how much our slope measurements change under this alternative approach.

\subsubsection{Slope changes in our scaling relations}
\label{sec:slope_changes_ND}
\begin{figure*}
    \centering
    \includegraphics[width = \textwidth]{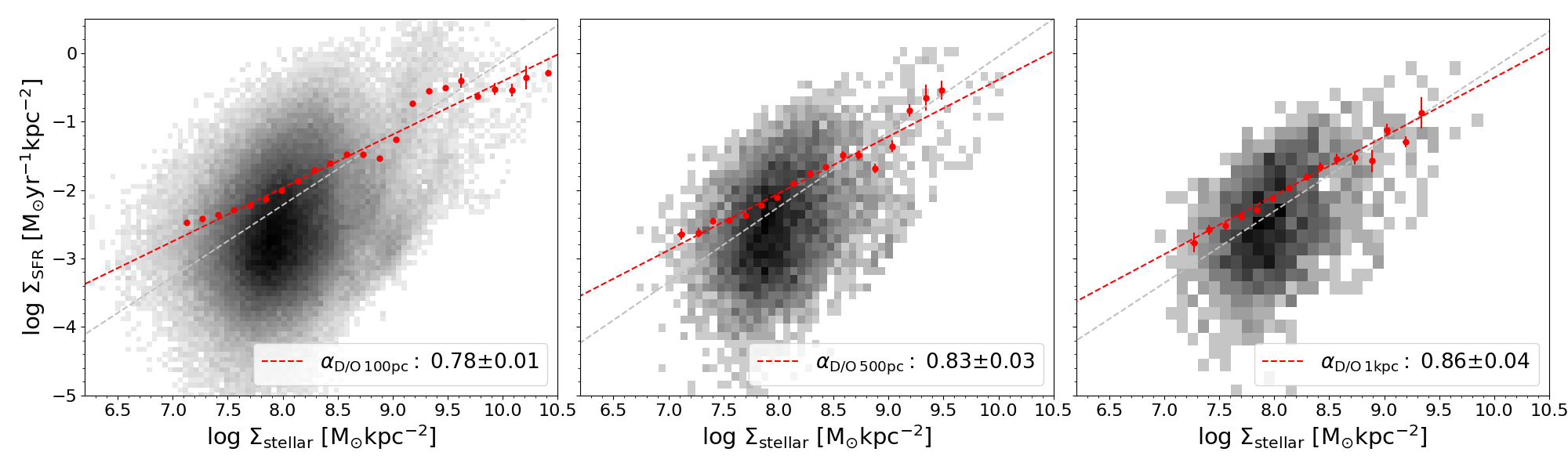}
    \caption{2D distribution of the resolved star formation main sequence considering all galaxies in our sample at three spatial resolutions ($100$~pc, $500$~pc and $1$~kpc from left to right). Non-detection pixels have been excluded from the fit. This produces a systematic flattening at higher spatial resolution. The fiducial overall best-fitting power law at each spatial scale is marked with the gray dashed line for reference.
    \label{fig:global_rSFMS_Hani}}
\end{figure*}

Figures~\ref{fig:global_rSFMS_Hani}, \ref{fig:global_KS_Hani} and~\ref{fig:global_MolGas_Hani}, show the rSFMS, rKS, and rMGMS, respectively, when non-detections are excluded from the fit. We define $\alpha_{\rm D/O}$ and $\sigma_{\rm D/O}$ as the measured slope and scatter considering pixels with nonzero detections only. Each relation studied here is presented at three spatial scales ($100$~pc, $500$~pc, and $1$~kpc from left to right).
The slopes and scatter measured in the overall sample at each spatial resolution are summarized in Table~\ref{tab:global_slopes_allres_Hani}, while Tables~\ref{tab:slopes_all_galaxies_rSFMS_Hani}, \ref{tab:slopes_all_galaxies_KS_Hani}, and~\ref{tab:slopes_all_galaxies_MGMS_Hani} list the slopes and scatter measured for the rSFMS, rKS, and rMGMS, respectively, in each galaxy at each spatial scale.
When we exclude non-detected pixels from our analysis, we see a systematic steepening of the slope at larger spatial scales in the rSFMS and the rMGMS. The steepening is particularly strong in the rMGMS due to lower (CO) detection fraction in this relation (${\sim}35\%$ at $100$ pc). The rKS is less affected as molecular gas and SFR share a  more similar spatial distribution \citep[][H.-A.~Pan et al.\ in prep.]{Schinnerer2019} and non-detections in SFR frequently coincide with a non-detection in molecular gas, resulting in a high detection fraction of SFR for a given molecular gas value in the rKS relation (${\sim}84\%$ at $100$ pc). 

Additionally, we observe a flattening at the low end of the stellar mass surface density  ($\log\Sigma_\mathrm{stellar}\ [\mathrm{M}_{\odot}~\mathrm{kpc}^{-2}] \lesssim 7.5$) in the rSFMS and the rMGMS at $100$~pc scale. This flattening is qualitatively similar to that reported by  \citet{Barrera-Ballesteros2021} or \citet{CanoDiaz2019}, attributed to H$\alpha$ pollution from non-SF emission at low SFR values and due to the H$\alpha$ detection threshold \citep[see also][]{Salim2007}. 
As we correct for non-star-forming contributions in our H$\alpha$ derived SFR maps, we interpret this flattening is caused by the larger fraction of non-detections in these low stellar mass surface density bins. Therefore, excluding non-detections leads to an overestimation of the mean measured in these bins, which results in the observed flattening at low stellar mass surface densities. 


Even though we see flatter relations at $100$~pc than at $1$~kpc for the overall sample when non-detections are excluded, the behavior of individual galaxies is more diverse. Differences between the lowest and highest resolution measurements in each galaxy mainly depend on the spatial distributions of stellar mass and ionized gas (in the case of the rSFMS) across the galactic disk within each galaxy. However, for all galaxies the slopes obtained at $100$~pc are flatter when non-detections are excluded, compared to when they are included.

\begin{table*}
\centering
\begin{tabular}{ccccccc}
\hline
\hline
Relation & $\alpha_{\mathrm{D/O}\,100\mathrm{pc}}$ & $\sigma_{\mathrm{D/O}\,100\mathrm{pc}}$ & $\alpha_{\mathrm{D/O}\,500\mathrm{pc}}$ & $\sigma_{\mathrm{D/O}\,500\mathrm{pc}}$ & $\alpha_{\mathrm{D/O}\,1\mathrm{kpc}}$ & $\sigma_{\mathrm{D/O}\,1\mathrm{kpc}}$ \\
\hline
rSFMS & $0.78\pm0.01$ & $0.49$ & $0.83\pm0.03$ & $0.47$ & $0.86\pm0.04$ & $0.44$ \\
rKS & $0.97\pm0.01$ & $0.41$ & $1.00\pm0.02$ & $0.33$ & $0.98\pm0.03$ & $0.27$ \\
rMGMS & $0.74\pm0.01$ & $0.32$ & $0.83\pm0.02$ & $0.30$ & $0.88\pm0.03$ & $0.29$ \\
\hline
\end{tabular}
\caption{Slope ($\alpha$) and scatter ($\sigma$) using all the available pixels in our sample, for each one of the scaling relations probed,  at spatial scales of $100$~pc, $500$~pc and $1$~kpc. Non-detections are excluded when measuring the slope.}
\label{tab:global_slopes_allres_Hani}
\end{table*}
\begin{table*}
\centering
\begin{tabular}{ccccccc}
\hline
\hline
Target & $\alpha_{\mathrm{D/O}\,100\mathrm{pc}}$ & $\sigma_{\mathrm{D/O}\,100\mathrm{pc}}$ & $\alpha_{\mathrm{D/O}\,500\mathrm{pc}}$ & $\sigma_{\mathrm{D/O}\,500\mathrm{pc}}$ & $\alpha_{\mathrm{D/O}\,1\mathrm{kpc}}$ & $\sigma_{\mathrm{D/O}\,1\mathrm{kpc}}$ \\
\hline
NGC1087 & $1.19\pm0.03$ & $0.41$ & $1.40\pm0.03$ & $0.35$ & $1.34\pm0.03$ & $0.23$ \\
NGC1300 & $0.57\pm0.01$ & $0.39$ & $0.74\pm0.01$ & $0.38$ & $0.71\pm0.01$ & $0.35$ \\
NGC1365 & $1.06\pm0.01$ & $0.50$ & $1.23\pm0.01$ & $0.53$ & $1.02\pm0.01$ & $0.43$ \\
NGC1385 & $1.33\pm0.03$ & $0.43$ & $1.42\pm0.03$ & $0.44$ & $1.44\pm0.03$ & $0.46$ \\
NGC1433 & $0.48\pm0.01$ & $0.36$ & $0.59\pm0.01$ & $0.33$ & $0.66\pm0.01$ & $0.26$ \\
NGC1512 & $0.35\pm0.01$ & $0.40$ & $0.63\pm0.01$ & $0.36$ & $0.78\pm0.01$ & $0.33$ \\
NGC1566 & $0.59\pm0.02$ & $0.48$ & $0.64\pm0.02$ & $0.46$ & $0.65\pm0.02$ & $0.41$ \\
NGC1672 & $1.04\pm0.01$ & $0.48$ & $1.10\pm0.01$ & $0.50$ & $1.18\pm0.01$ & $0.50$ \\
NGC2835 & $0.39\pm0.03$ & $0.41$ & $0.51\pm0.03$ & $0.36$ & $0.60\pm0.03$ & $0.22$ \\
NGC3351 & $0.40\pm0.02$ & $0.37$ & $1.14\pm0.02$ & $0.28$ & $1.24\pm0.02$ & $0.13$ \\
NGC3627 & $0.23\pm0.03$ & $0.48$ & $0.32\pm0.03$ & $0.42$ & $0.24\pm0.03$ & $0.40$ \\
NGC4254 & $0.62\pm0.01$ & $0.48$ & $0.64\pm0.01$ & $0.41$ & $0.72\pm0.01$ & $0.28$ \\
NGC4303 & $0.53\pm0.01$ & $0.50$ & $0.66\pm0.01$ & $0.42$ & $0.51\pm0.01$ & $0.38$ \\
NGC4321 & $0.71\pm0.01$ & $0.49$ & $0.86\pm0.01$ & $0.49$ & $0.95\pm0.01$ & $0.45$ \\
NGC4535 & $0.35\pm0.03$ & $0.49$ & $0.32\pm0.03$ & $0.53$ & $0.54\pm0.03$ & $0.41$ \\
NGC5068 & $0.11\pm0.04$ & $0.46$ & $0.57\pm0.04$ & $0.33$ & $-0.48\pm0.04$ & $0.24$ \\
NGC7496 & $0.77\pm0.03$ & $0.46$ & $1.25\pm0.03$ & $0.46$ & $1.19\pm0.03$ & $0.36$ \\
IC5332 & $0.20\pm0.05$ & $0.36$ & $0.12\pm0.05$ & $0.25$ & $-0.09\pm0.05$ & $0.06$ \\
\hline
\end{tabular}
\caption{Slope ($\alpha$) and scatter ($\sigma$) for the rSFMS measured in each galaxy in our sample at spatial scales of $100$~pc, $500$~pc and $1$~kpc. The non-detections have been excluded from the slope measurement. }
\label{tab:slopes_all_galaxies_rSFMS_Hani}
\end{table*}

\subsubsection{Treatment of non-detections and its effect on the slope of the scaling relations}
\label{sec:quantification_slope_change_ND}

As explained in Sec.~\ref{sec:steepening}, the flattening of the relations toward small spatial scales arises when non-detections are excluded from the analysis. This is because ``empty'' regions at smaller spatial scales are averaged at larger spatial scales with regions including detections. This effectively dilutes the signal when the data are degraded, and the magnitude of this dilution will depend on the local detection fraction.
The spatial distribution of H$\alpha$ with respect to the stellar mass surface density is complex, and varies not only from galaxy to galaxy, but also between different environments within galaxies. Therefore, the local detection fraction follows an equally complex distribution.

\begin{figure}
    \centering
    \includegraphics[width = \columnwidth]{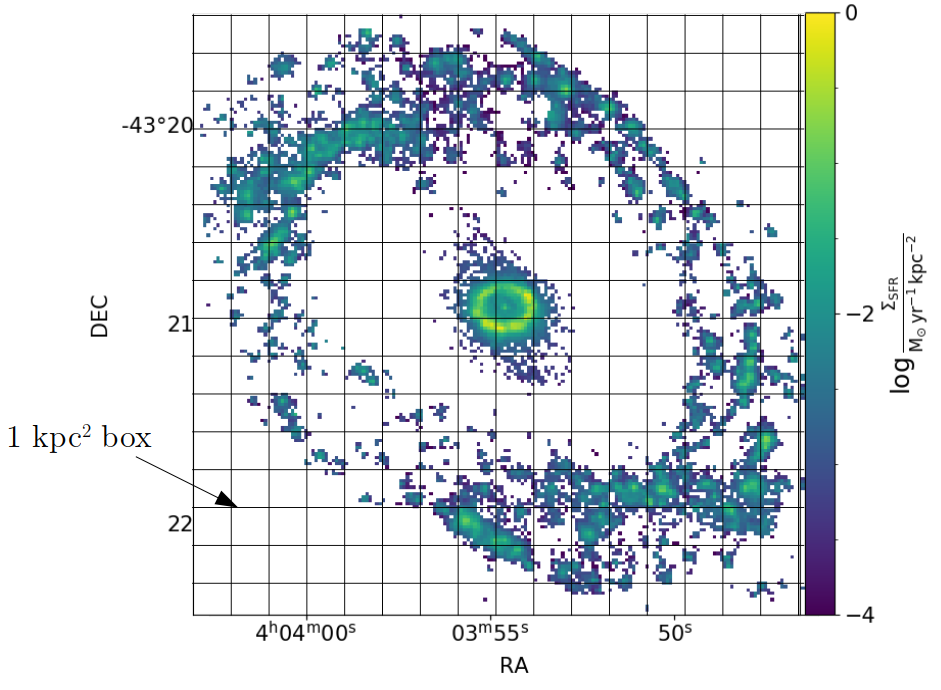}
    \caption{Sketch to exemplify the methodology to compute the distribution of the H$\alpha$ detection fraction. In each 1~kpc$^{2}$ box we calculated the filling factor ($\mathrm{DF}_{\mathrm{H}\alpha}$) defined as the fraction of pixels with nonzero SFR.
    \label{fig:grid_example}}
\end{figure}

Here we explore if we can recover a relation between the detection fraction and the amount of flattening, despite of the complex distribution of these quantities, using the slope measured at $100$~pc spatial scale in a given galaxy, when N/D are excluded from the measurement. We measured the detection fraction of the SFR inside $1$~kpc$^2$ boxes in the $100$~pc spatial scale maps. For each box we calculated the corresponding detection fraction ($\mathrm{DF}_{\mathrm{H}\alpha}$), defined as the fraction of pixels with nonzero emission, and we computed the distribution of the  $\mathrm{DF}_{\mathrm{H}\alpha}$ within these boxes for each galaxy. Figure~\ref{fig:grid_example} shows a schematic representation of the approach for one example galaxy.

\begin{figure*}
    \centering
    \includegraphics[width = 0.85\textwidth]{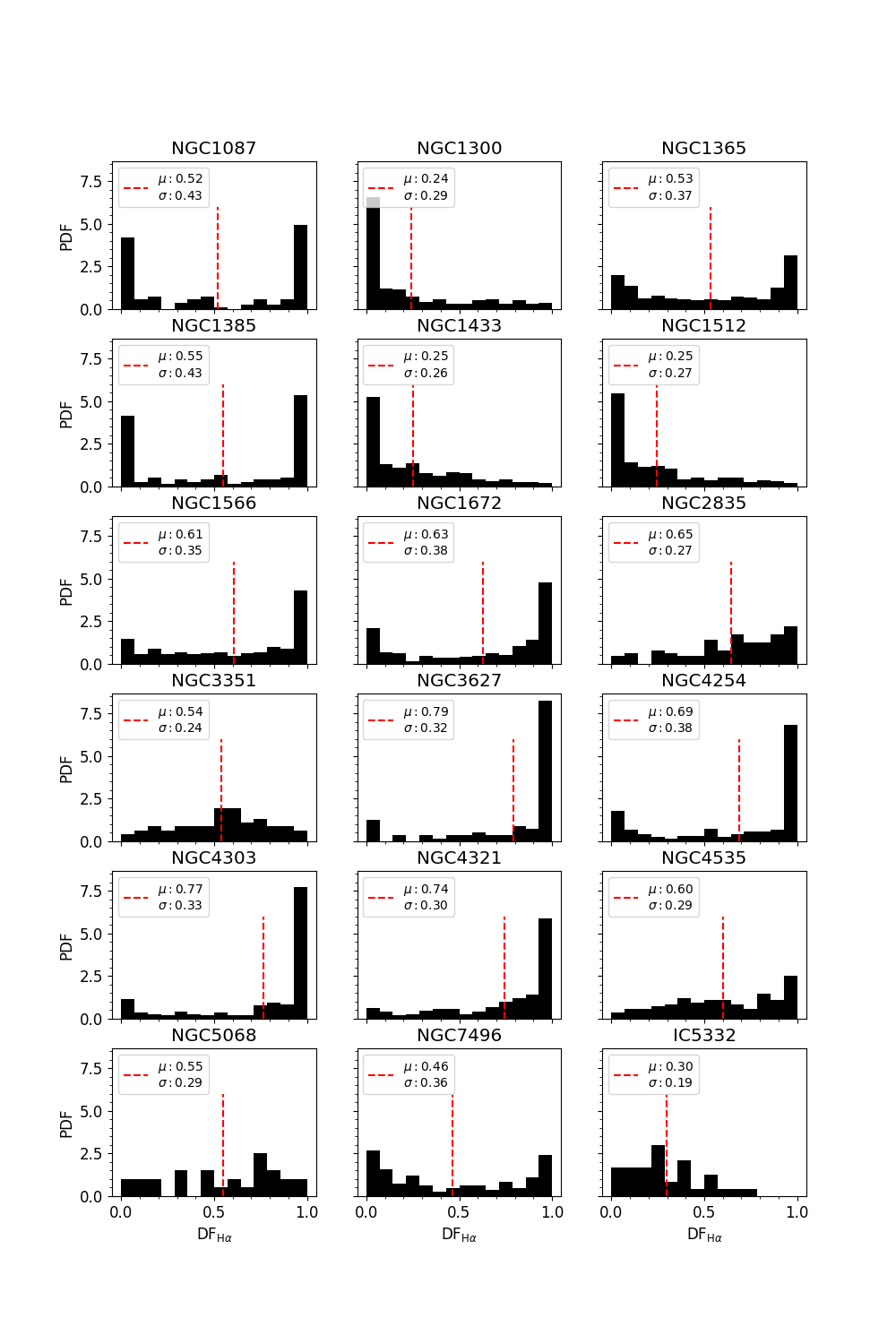}
    \caption{Normalized distribution of the H$_{\alpha}$ detection fraction ($\mathrm{DF}_{\mathrm{H}\alpha}$) inside 1~kpc$^2$ size boxes in the galactic disk. Each panel corresponds to a galaxy from our sample. The mean ($\mu$) and standard deviation ($\sigma$) are indicated in each panel.
    \label{fig:Ha_ff_dist}}
\end{figure*}

Figure~\ref{fig:Ha_ff_dist} shows the normalized distribution of $\mathrm{DF}_\mathrm{H\alpha}$ for each galaxy. The corresponding mean ($\mu_{\mathrm{DF}}$) and standard deviation ({$\sigma_{\mathrm{DF}}$}) are indicated in each panel. The latter contains information about the spatial configuration of the SFR. A galaxy with a compact (spatially concentrated) configuration will have a rather bimodal distribution of $\mathrm{DF}_\mathrm{H\alpha}$ (and consequently higher $\sigma_{\mathrm{DF}_\mathrm{H\alpha}}$), where boxes will have mostly ${\sim}0$ or ${\sim}1$ values, whereas a more clumpy (assembled as individual separated clumps) configuration will lead to a flatter and uniform distribution. 

We parametrize the difference in slope when non-detections are excluded from the fit as compared to the fiducial value as:
\begin{equation}
    \Delta\alpha_\mathrm{D/O} = \alpha - \alpha_\mathrm{D/O}~,
\end{equation}
where $\alpha$ is the slope measured for a galaxy using the fiducial approach and $\alpha_\mathrm{D/O}$ is the slope measured for the same galaxy when non-detections are excluded in the fit.

\begin{figure}
    \centering
    \includegraphics[width = \columnwidth]{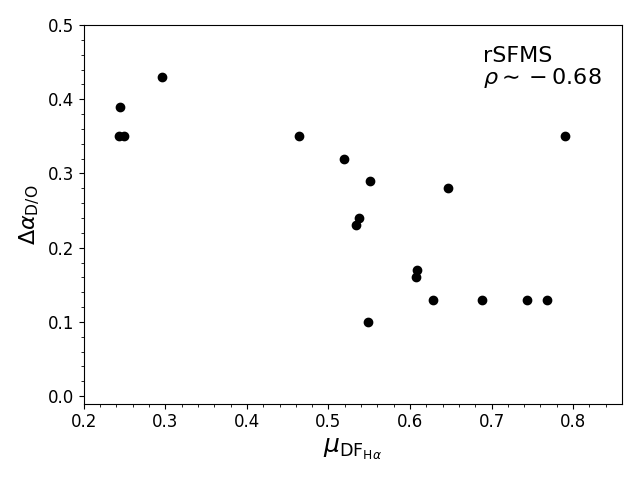}
    \caption{Difference between the slope of the rSFMS of each galaxy measured with the fiducial approach and the slope measured when non-detections are excluded, as a function of the mean detection fraction of the SFR tracer in each galaxy ($\mu_{\mathrm{DF}_{\mathrm{H}\alpha}}$). All values are positive in the $y$-axis because excluding non-detections always leads to a flatter relation. The PCC of the correlation is indicated in the top-right corner of the panel.}
    \label{fig:Ha_ff_vs_slope}
\end{figure}

Figure~\ref{fig:Ha_ff_vs_slope} shows the change of slope when non-detections are excluded in the calculation as a function of  $\mu_{\mathrm{DF}_\mathrm{H\alpha}}$ for the galaxies in our sample. 
The figure shows a negative correlation ($\mathrm{PCC} \approx {-}0.68$) between $\mu_{\mathrm{DF}_\mathrm{H\alpha}}$ and the change in the measured slope when non-detections are excluded. This can be interpreted as galaxies with higher average detection fraction are less affected by dilution of their signal, which implies that excluding non-detections has a smaller impact on the measured slope. 
So, despite of the complexity of the 2D distribution of ionized gas across the galactic disk, we identify a trend between the flattening and the distribution of the detection fraction.

The scatter in the correlation shown in Figure~\ref{fig:Ha_ff_vs_slope} is likely because this diagnostic does not include any aspect of the underlying $\Sigma_\mathrm{stellar}$ (or $\Sigma_\mathrm{mol. gas}$) distribution. Ultimately, the change of the slope will not only depend on the detection fraction in a given region, but also on its underlying $\Sigma_\mathrm{stellar}$ (for the rSFMS). Whereas the first is highly driven by local environment \citep[][Meidt et al. submitted]{Querejeta2021}, the second varies much more smoothly across galactic disks. The detection fraction distribution by its own cannot completely describe the flattening of the slope.


\subsubsection{Role of detection fraction in the dispersion of slopes at 100~pc}
\label{sec:dispersion_slopes}

\begin{figure}
    \centering
    \includegraphics[width = \columnwidth]{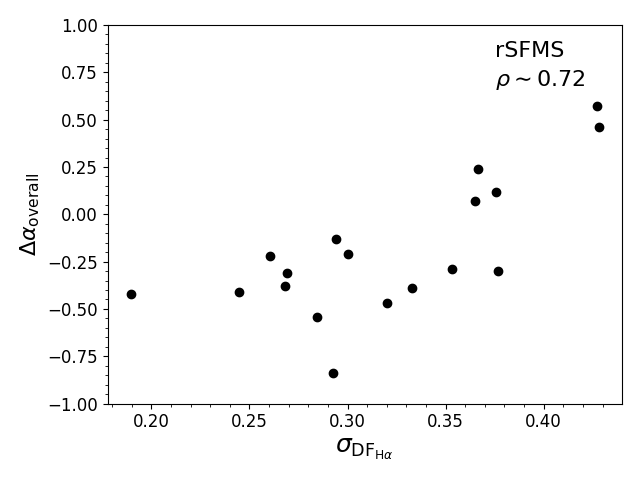}
    \caption{Difference between the slope of the rSFMS of each galaxy measured with the fiducial approach and the overall measurement as a function of the standard deviation of the detection fraction distribution of the SFR tracer in each galaxy ($\sigma_{\mathrm{DF}_{\mathrm{H}\alpha}}$). The PCC of the correlation is indicated in the top-right corner of the panel.}
    \label{fig:Ha_ff_vs_slope_global}
\end{figure}

In this section, we test if the dispersion that we see in the slopes measured for the rSFMS could be linked to differences in the SFR detection fraction distributions.
Figure~\ref{fig:Ha_ff_vs_slope_global} shows $\Delta\alpha_\mathrm{overall}$ (as defined in Eq.~\ref{eq:global_def}, the difference in slope of a given galaxy and the full sample) as a function of $\sigma_{\mathrm{DF}_\mathrm{H\alpha}}$. The figure shows a clear positive correlation ($\mathrm{PCC} \approx 0.72$), meaning that galaxies with larger spread (i.e., larger $\sigma_{\mathrm{DF}_\mathrm{H\alpha}}$) tend to have a steeper rSFMS (i.e., more positive $\Delta\alpha_\mathrm{overall}$) and galaxies with smaller spread tend to have a flatter rSFMS (i.e., more negative $\Delta\alpha_\mathrm{overall}$). 
This can be explained as galaxies with larger $\sigma_{\mathrm{DF}_\mathrm{H\alpha}}$ (i.e., with a more bimodal distribution of $\mathrm{DF}_\mathrm{H\alpha}$) are associated with a spatial configuration in which the impact of non-detections is stronger in some regions (usually in the low stellar mass surface density regime) and less significant in others (central regions with higher stellar mass surface densities). This leads to a systematically steeper rSFMS by pulling down the low mass surface density end of the rSFMS, where non-detections dominate, with respect to the high mass surface density end. Galaxies with smaller $\sigma_{\mathrm{DF}_\mathrm{H\alpha}}$ values show either a flatter $\mathrm{DF}_\mathrm{H\alpha}$ distribution or a concentration of their values around zero. In these scenarios, non-detections do not contribute to steepen the relation in the same way, since different mass ranges are similarly affected. 

We find similar correlations between the detection fraction distributions of the molecular gas tracer and changes in the slope of the rMGMS (see Appendix~\ref{sec:ap_DF_rMGMS}). Hence, the detection fraction of H$\alpha$ (or molecular gas) is a relevant aspect to set the rSFMS (rMGMS) slope when measured at high resolution. This is also consistent with finding a similar dispersion in the slopes of the rSFMS and the rMGMS, and a significantly smaller dispersion in the slopes of the rKS. Due to the more similar spatial distribution between H$\alpha$ and the molecular gas tracer, differences in the detection fraction are also less extreme.

\subsection{Role of baryonic mass surface density in regulating SFR}
\label{sec:baryonic_vs_SFR}
\begin{figure}
    \includegraphics[width = \columnwidth]{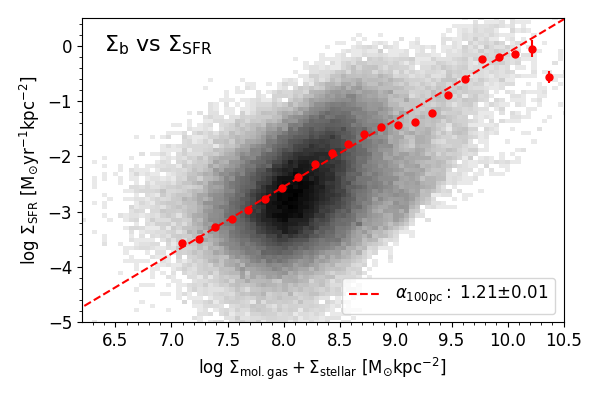}
    \caption{$\Sigma_\mathrm{SFR}$ as a function of $\Sigma_\mathrm{b}$, defined as $\Sigma_\mathrm{mol. gas} + \Sigma_\mathrm{stellar}$. We explored the existence of a tighter correlation with $\Sigma_\mathrm{b}$ as it is a better tracer for the total hydrostatic pressure exerted by the baryonic mass. We measured a slope of  $1.18\pm0.005$ with a scatter, and a scatter lower than the rSFMS and higher than the rKS.}
    \label{fig:Sigma_bar}
\end{figure}

Previous studies have demonstrated that the mid-plane pressure of the interstellar medium impacts its physical properties and consequently the local SFR surface density.
\citet{Leroy2008} reported a good correlation between the mid-plane hydrostatic gas pressure and the local SFR surface density and gas depletion time in 23 nearby galaxies in the THINGS \citep{Walter2008} and HERACLES \citep{Leroy2009} surveys. Along the same line, \citet{Schruba2019} --studying 8 local galaxies-- and \citet{Sun2020b} --studying 28 star-forming galaxies from the PHANGS-ALMA sample-- found that the average internal pressure of molecular clouds tends to balance the sum of their own weights and the external interstellar medium pressure. \citet{Sun2020b} also reported a tight relationship between the mid-plane hydrostatic pressure and SFR surface density across their sample.
These observations are in line with feedback-driven star formation models \citep{Ostriker2010,Ostriker2011}, in which the dynamical equilibrium of the interstellar medium in the galaxy gravitational potential regulates the local SFR surface density and vice versa.

Since both stars and molecular gas define the gravitational potential of a galaxy (neglecting the contribution from atomic gas), previous studies have also explored the existence of a relation between $\Sigma_\mathrm{SFR}$ and a combination of $\Sigma_\mathrm{stellar}$ and $\Sigma_\mathrm{mol. gas}$ \citep{Matteucci1989, Shi2011, Dib2017, Shi2018, Dey2019, Lin2019, Barrera-Ballesteros2021, Sanchez2021}.

In \citet{Barrera-Ballesteros2021}, the authors computed the ``baryonic'' mass surface density as $\Sigma_\mathrm{b} = \Sigma_\mathrm{stellar} + \Sigma_\mathrm{mol. gas}$ at $\sim$kpc scales and found that $\Sigma_\mathrm{b}$ tightly correlates with SFR, with a slope of $0.97$ and a residual scatter lower than that measured for the rSFMS or the rKS.
We derived the $\Sigma_\mathrm{b}$ versus $\Sigma_\mathrm{SFR}$ relation at $100$~pc spatial scale (see Fig.~\ref{fig:Sigma_bar}) and measured a slope of $1.21\pm0.01$ and a scatter of $0.49$~dex. At $500$~pc we obtain a slope of $1.22\pm0.02$ and a scatter of $0.46$~dex, and at $1$~kpc a slope of $1.14\pm0.03$ and a scatter of $0.42$~dex. These values  of scatter are higher than computed for the rKS and lower than those of the rSFMS at all spatial scales.  Thus, we find that $\Sigma_\mathrm{mol. gas}$ is a better predictor of $\Sigma_\mathrm{SFR}$ than $\Sigma_\mathrm{b}$. Differences with \citet{Barrera-Ballesteros2021} are probably related to the treatment of N/D in the fit (which may directly impact the measured slope, as shown in Sec.\ref{sec:steepening}), and the use of  Balmer decrement as molecular gas surface density tracer (which is expected to introduce some level of scatter to the correlation).

However, we stress here that the mid-plane hydrostatic pressure is not necessarily directly proportional to the baryonic mass surface density. Although $\Sigma_\mathrm{b}$ is sometimes used as a proxy for the hydrostatic pressure, it does not accurately reflect the gravitational potential felt by the gas disk in galaxies, because the stellar disk and gas disk usually have quite different scale heights \citep[e.g., ][]{Ostriker2010, Bacchini2019a, Bacchini2019b, Bacchini2020, Sun2020}. More careful investigations of the quantitative relation between $\Sigma_\mathrm{mol}$, $\Sigma_\mathrm{SFR}$, and the mid-plane hydrostatic pressure in the future will shed more light on the role of hydrostatic pressure in regulating star formation \citep[e.g., ][]{Barrera-Ballesteros2021, Barrera-Ballesteros2021b}.

\subsection{Systematic effects}

In this section, we test how some of our assumptions impact our measurements of the scaling relations probed. In Sec.~\ref{sec:aperture}, we derive the rSFMS and rKS relations after removing diffuse ionized gas (DIG) emission. In Sec.~\ref{sec:conv_fact}, we use a constant $\alpha_\mathrm{CO}$ to measure the rKS and the rMGMS. Finally, in Sec.~\ref{sec:SFR_from_SFH} we show the resultant rSFMS and rKS when a longer timescale SFR tracer is used instead of H$\alpha$.




\subsubsection{Role of diffuse gas emission}
\label{sec:aperture}

In our fiducial approach, we aim at measuring scaling relations consistently at different spatial scales. Thus, we want to include any detectable SFR emission at high angular resolution (as well as N/D), that would then be averaged in the lower resolution measurement. However, even though we followed the approach described in Sec.~\ref{sec:SFR} in order to correct the H$\alpha$ emission by DIG contamination, we could still be left with some amounts of contaminant emission in our analysis.
Here, we explore an alternative way to identify the star formation-associated  H$\alpha$ emission at the native MUSE resolution that is suitable to identify individual \hii\ regions.
For this, we use the \hii\ region catalog from F.~Santoro et al.\ (in prep.). In short, \hii\ regions are identified in the H$\alpha$ map using HIIphot \citep{Thilker2000} for resolved sources and DAOStarFinder \citep{Bradley2020} for point-like sources. A BPT diagnosis using the [\oiii]/H$\beta$ and [\nii]/H$\alpha$ line ratios of the integrated spectra of each region was then used to select the H$\alpha$-emitting regions dominated by star formation, as described in \citet{Kewley2006}. Once \hii\ regions have been identified at the native MUSE resolution, we mask all remaining emission before computing our maps at $100$~pc, $500$~pc, and $1$~kpc scales. The fraction of emission removed is on the order of ${\sim}30\%$ (${\sim}60\%$ of the pixels with nonzero emission in the fiducial approach are masked), consistent with the diffuse fractions obtained by decomposing the narrowband H$\alpha$ images of these galaxies into diffuse and compact components in Fourier space \citep{Hygate2019, Chevance2020b}. This demonstrates that our procedure reliably removes all potentially contaminant emission before resampling the SFR map.

\begin{figure}
    \centering
    \includegraphics[width = \columnwidth]{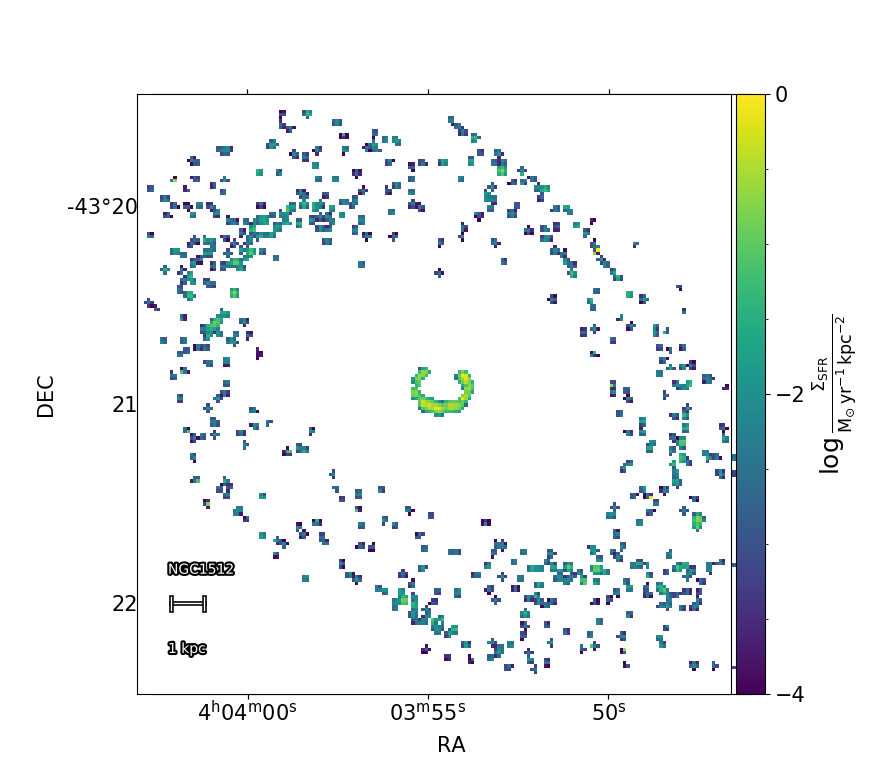}b
    \caption{Example of SFR surface density at spatial scales of $100$~pc, for one of the galaxies in our sample (NGC~1512). All the H$\alpha$ emission not associated with morphologically-defined \hii\ regions has been excluded before re-sampling the maps. The methodology used to exclude this emission is described on the main text in Sec.~\ref{sec:aperture}.
    \label{fig:mapsNGC1512_HIIreg}}
\end{figure}

\begin{figure}
    \centering
    \includegraphics[width = \columnwidth]{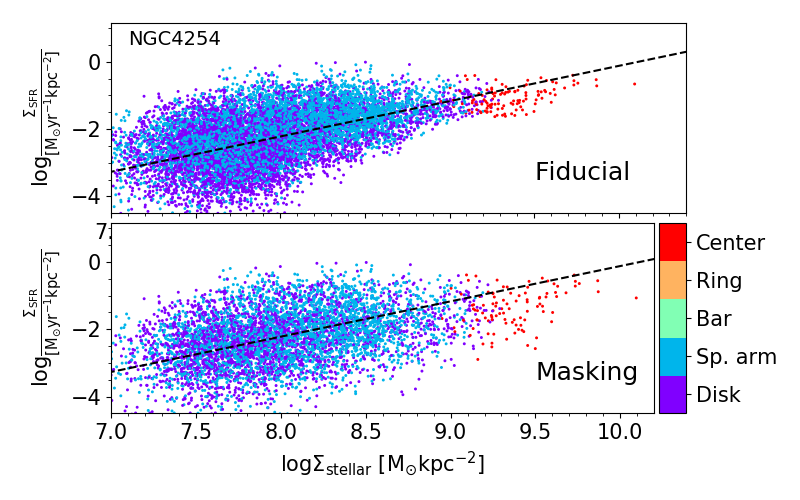}
    \caption{rSFMS of one galaxy in our sample (NGC~4254) to illustrate the effect of removing H$\alpha$ emission not associated with morphologically defined \hii\ regions before resampling the SFR surface density map. Top: fiducial methodology. Bottom: alternative methodology as described in Sec.~\ref{sec:aperture}. The color scheme is the same as in Fig.~\ref{fig:KS_app100}. The black line correspond to the global rSFMS measured using all pixels with the fiducial approach (see Fig.~\ref{fig:global_all}). }
    \label{fig:example_aperture}
\end{figure}

Figure~\ref{fig:mapsNGC1512_HIIreg} shows the resulting SFR map from this different approach at a spatial scale of $100$~pc for NGC~1512. Individual \hii\ regions are much more easily distinguished in the resultant map. Quantitatively, this creates large differences, mostly at the edges of \hii\ regions or molecular clouds, where the signal is now averaged with neighboring zero-emission pixels during the resampling and pulled to lower levels of SFR. At the same time, it removes all faint, DIG-contaminated pixels from the relation. Figure~\ref{fig:example_aperture} highlights this effect for the rSFMS of NGC~4254. The top panel shows the rSFMS from the fiducial approach, and the bottom panel shows the same relation using this alternative approach. The color scheme is the same as used in Fig.~\ref{fig:rSFMS_app100}, indicating the morphological environment of each pixel. The black dashed line corresponds to the overall fit from the fiducial approach and has been overplotted as reference. Faint DIG-contaminated pixels are removed, which reduces the number of data points in the relation. However, the remaining pixels in the low SFR regime are pulled to even lower values resulting in a qualitatively similar rSFMS.

\begin{figure*}
    \centering
    \includegraphics[width = \textwidth]{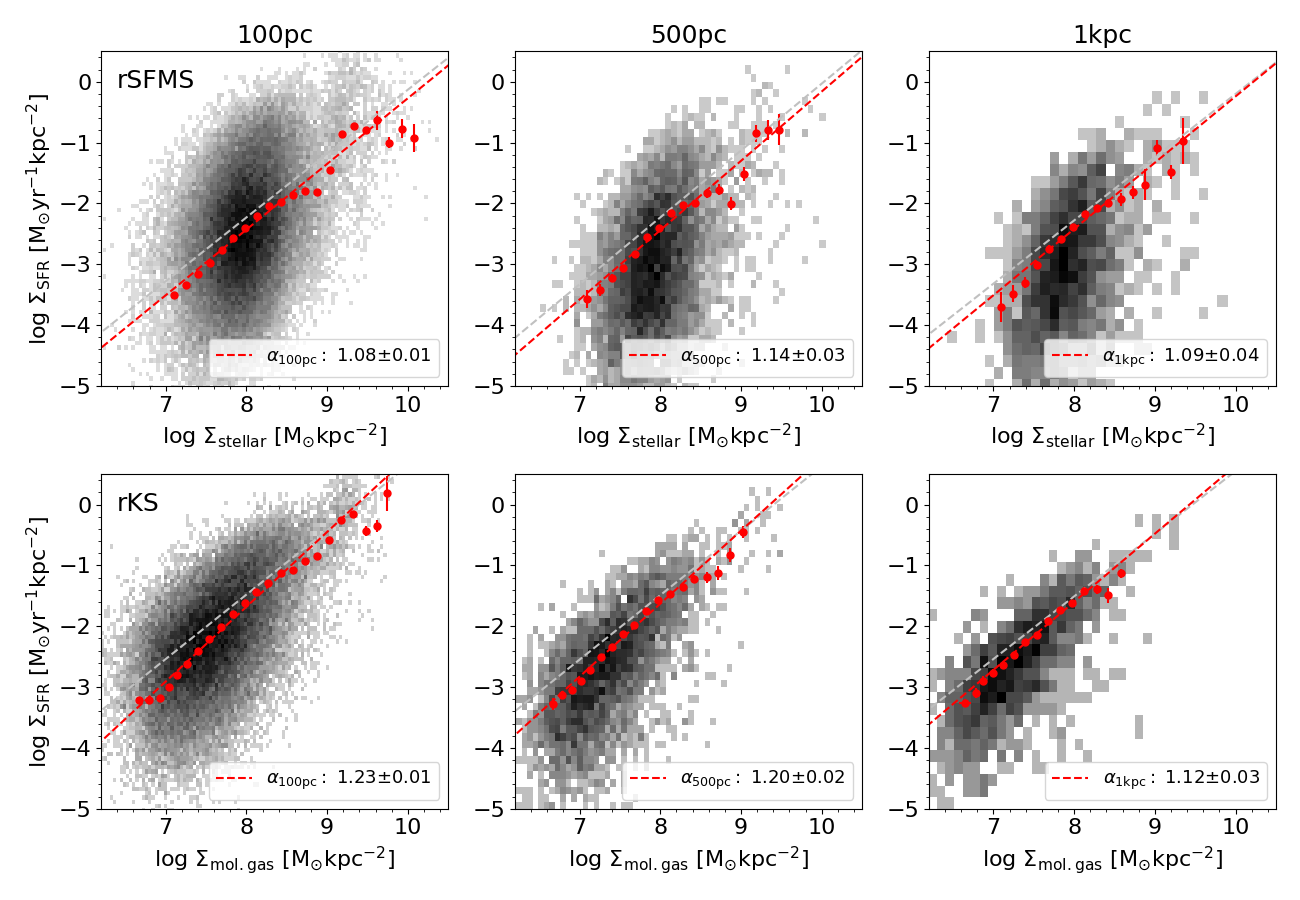}
    \caption{Effect of excluding diffuse gas emission before resampling: 2D distributions of the overall rSFMS (top row) and rKS (bottom row) using all the available pixels in our sample, probed at spatial scales of $100$~pc (left column), $500$~pc (middle column) and $1$~kpc (right column). The $x$-axis binning and the best-fitting power law are indicated with red dots and a red dashed line respectively. The measured slope and its error are indicated in each panel (see Sec.~\ref{sec:aperture} for details). The fiducial overall best-fitting power law at each spatial scale is marked with the gray dashed line for reference.}
    \label{fig:global_variation_scale_apert}
\end{figure*}

Figure~\ref{fig:global_variation_scale_apert} shows the 2D distributions for the rSFMS and the rKS probed at $100$~pc, $500$~pc, and $1$~kpc scale with their corresponding binned trends and best-fitting power laws.
The inferred slopes agree relatively well with those obtained using the fiducial method (see Fig.~\ref{fig:variation_scale}). Differences are expected since we are now fitting a power law to a subset of pixels (while the rest is classified as N/D). However, the scatter is systematically enhanced in both relations by a factor of $\sim1.1 - 1.4$, depending on the spatial scale considered. This increase is mainly driven by the effect previously described: the dilution of emission at the edges of \hii\ regions or molecular clouds. 
Thus, we conclude that we could be underestimating the SF-associated emission by removing too much signal and, thus, introducing additional scatter to the relation. 
The true scatter of these scaling relations is probably somewhere in between our fiducial approach and applying the strict \hii\ region mask. Hence, in our fiducial approach the scatter of these relations is possibly underestimated due to contaminant H$\alpha$ emission boosting the SFR of the faintest pixels.


\begin{table*}
\centering
\begin{tabular}{ccccccc}
\hline
\hline
Relation & $\alpha_{100\mathrm{pc}}$ & $\sigma_{100\mathrm{pc}}$ & $\alpha_{500\mathrm{pc}}$ & $\sigma_{500\mathrm{pc}}$ & $\alpha_{1\mathrm{kpc}}$ & $\sigma_{1\mathrm{kpc}}$ \\
\hline
rSFMS & $1.08\pm0.01$ & $0.58$ & $1.14\pm0.03$ & $0.63$ & $1.09\pm0.04$ & $0.61$ \\
rKS & $1.23\pm0.01$ & $0.48$ & $1.20\pm0.02$ & $0.42$ & $1.12\pm0.03$ & $0.35$ \\
\hline
\end{tabular}
\caption{Slope ($\alpha$) and scatter ($\sigma$) using all available pixels in our sample, for the scaling relations probed,  at spatial scales of $100$~pc, $500$~pc and $1$~kpc. All the H$\alpha$ emission not associated with morphologically-defined \hii\ regions has been excluded before re-sampling the maps to the different spatial scales. The methodology used to exclude this emission is described in Sec.~\ref{sec:aperture}}
\label{tab:global_slopes_allres_apert}
\end{table*}

\subsubsection{Impact of the chosen CO-to-\texorpdfstring{H$_2$}{H2} conversion factor}
\label{sec:conv_fact}

As explained in Sec.~\ref{sec:alma}, our fiducial CO-to-H$_{2}$ conversion factor scales with local gas-phase metallicity. In this section, we explore the impact of assuming a constant conversion factor of $4.35$ M$_{\odot}$~pc$^{-2}$ (K~km~s$^{-1}$)$^{-1}$  \citep{Bolatto2013} on the measured rKS and rMGMS. 

\begin{figure}
    \includegraphics[width = \columnwidth]{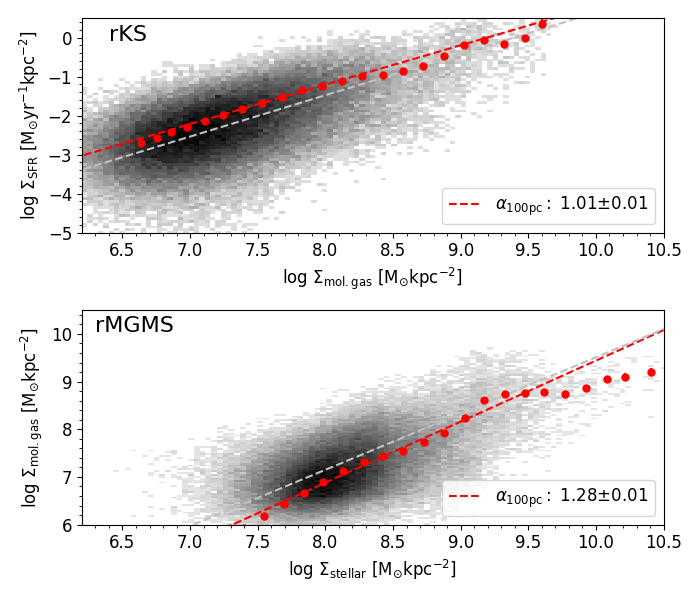}
    
    \caption{rKS (top) and rMGMS (bottom) measured using all pixels from our sample and assuming a constant CO-to-H$_{2}$ conversion factor. This conversion factor leads to a flatter rKS and a steeper rMGMS (see Sec.~\ref{sec:conv_fact} for a discussion). The best fit power law obtained with the fiducial approach is indicated with the gray line.}
    \label{fig:constantCO}
\end{figure}

All of the galaxies in our sample exhibit either a negative or flat metallicity profile \citep{Ho2015}. This leads to a higher CO-to-H$_{2}$ conversion factor in the outer part of the galactic disk than in the central part.
When the radial gradient is removed, differences between central (usually denser) clouds and outer  (usually fainter) clouds are enhanced. This widens the range of molecular gas surface densities probed.
Figure~\ref{fig:constantCO} shows the rKS (top) and rMGMS (bottom) at $100$~pc spatial scale using a constant $\alpha_\mathrm{CO}$. 
The best-fitting power law obtained with the fiducial approach is indicated by the gray line.
As a consequence of this widened range of molecular gas surface densities from applying a constant CO-to-H$_{2}$ conversion factor, the slope of the rKS decreases to $1.01\pm0.01$ and that of the rMGMS increases to $1.28\pm0.01$, representing a change of ${\sim}5\%$ and ${\sim}8\%$, respectively, as compared to the fiducial scenario. Finally, the normalization of these relations is also affected by the CO-to-H$_{2}$ prescription. Under the assumption of a constant $\alpha_\mathrm{CO}$, we find an intercept of $-9.29$ for the rKS and $-3.36$ for the rMGMS (for reference, the intercepts computed under our fiducial $\alpha_\mathrm{CO}$ prescriptions are $-9.96$ and $-2.23$, respectively).

\subsubsection{Probing SFR on a 150~Myr timescale with spectral fitting}
\label{sec:SFR_from_SFH}

So far, we have used H$\alpha$ emission as our SFR tracer. H$\alpha$ is known to trace the formation of stars during the last ${\sim}10$~Myr \citep{Calzetti-book, Leroy2012, CatalanTorrecilla2015, Haydon2020}. Here we explore how the slope measured for the rSFMS and rKS varies when we use a SFR tracer sensitive to longer star formation timescales. We use the resolved star formation histories (see Sec.~\ref{sec:SSP_fit}) to map the fraction of total stellar mass assembled during the last $150$~Myr (i.e., the youngest 4~age bins in our age--metallicity grid). Multiplying this fraction with the stellar mass surface density and dividing it by $1.5\times10^{7}$~yr results in a SFR surface density map that probes longer timescales.

\begin{figure}
    \includegraphics[width = \columnwidth]{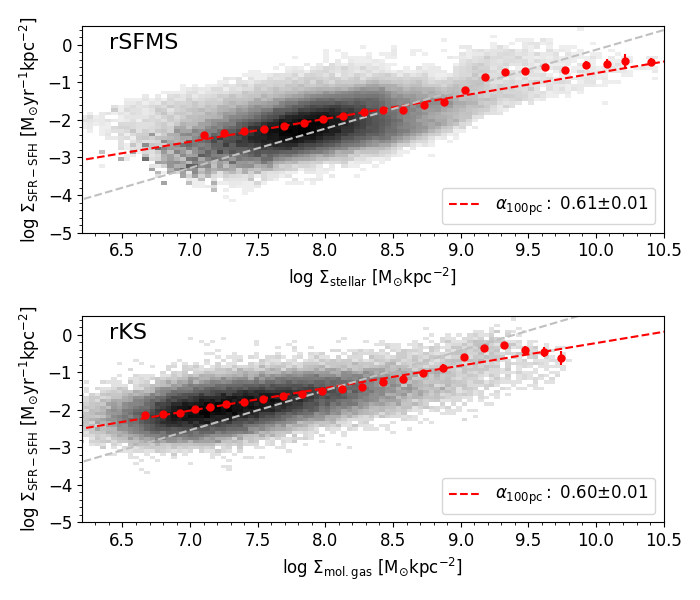}
    
    \caption{rSFMS (top) and rKS (bottom) measured at $100$~pc resolution, using all available pixels from our sample and adopting a SFR tracer with a longer timescale. We have used the derived SFH to compute a SFR tracer sensitive to SF episodes in the last $150$~Myr. This smooths the SFR measurements, flattens both relations, and reduces their scatter. The red dashed line show best-fitting power law to the binned data (red points). The best fit obtained with the fiducial approach is indicated with the gray line.}
    \label{fig:SFR_from_SFH}
\end{figure}

A longer timescale SFR tracer smooths the SFR values, i.e., high SFR values from the short timescale are nearly unaffected and low SFR values are pushed to higher values. Figure~\ref{fig:SFR_from_SFH} shows how this change impacts the rSFMS (top) and the rKS (bottom). 
Consequently, both relations are significantly flattened. In particular, the slope of the rSFMS decreases to $0.61\pm0.01$ and the slope of the rKS to $0.60\pm0.01$. 

This also drastically reduces the scatter in both relations to $0.27$~dex and $0.24$~dex, respectively.  A decrease in the scatter is consistent with the uncertainty principle reported in \citet{Kruijssen2014} and \cite{Kruijssen2018} (also see discussion in Sec.~\ref{sec:main_findings}). Averaging the SFR over a longer time scale increases the period of time in which young stars can be detected (i.e., $\tau_\mathrm{young-stars} > \tau_{\mathrm{H}\alpha}$), and thus reduces the critical spatial scale at which the relation breaks due to statistical undersampling of the star formation process. 

We summarize that the timescales involved in the SFR determination strongly influence the resulting slope and scatter of the rSFMS and rKS relations. This agrees with a similar finding presented in \citet{Barrera-Ballesteros2021}.

\subsection{Implication of our results}
\label{sec:imply}

While the relation between $\Sigma_\mathrm{SFR}$ and $\Sigma_\mathrm{mol. gas}$ is direct, since star formation occurs in molecular clouds, the relation between $\Sigma_\mathrm{SFR}$ and $\Sigma_\mathrm{stellar}$ is understood as the interplay between the hydrostatic pressure exerted by the stellar (and cold gas) disk, together with feedback processes (such as stellar winds and supernovae) regulating the star formation. However, given that we found the rKS was tighter than the rSFMS and the $\Sigma_\mathrm{SFR}$ versus $\Sigma_{\mathrm{b}}$ correlation, we conclude that it is mainly the amount of available molecular gas that regulates the star formation rate rather than the amount of stellar or baryonic mass. This agrees well with what was reported by \citet{Lin2019}, where the authors also conclude that the rSFMS could originate from the existence of the rKS and the rMGMS, where the latter can be explained either as the molecular gas following the gravitational potential of the stellar disk or both, stars and gas following the same underlying potential defined by the total mass. The similarities between the rSFMS and the rMGMS in our data across different galactic environments (see Figs.~\ref{fig:rSFMS_app100} and~\ref{fig:MolGas_app100}) and the high scatter of the rSFMS as compared to the other two scaling relations suggest that this might be the case, and that the substantial variations in the rSFMS are driven by a combination of abundance/\linebreak[0]{}lack~of molecular gas to fuel star formation as well as variations in star formation efficiency.

The rMGMS is the relation with the least scatter among the three relations at a spatial scale of $100$~pc, consistent with the expectation from the perpective of the time-scales of the star-forming cycle. The same was recently reported in \citet{Ellison2020} in an analysis carried out at kpc spatial scales. Interestingly, the rKS exhibits the most homogeneous behavior across different environments and between galaxies when measured at $100$~pc scales. Variations in the slope of the rSFMS and the rMGMS across different galaxies seem to be related to differences in the detection fraction of either the SFR tracer or the molecular gas tracer. 
This effect is less important in the rKS, since molecular and ionized gas share more similar spatial distributions.


\section{Summary}
\label{sec:summary}

We have used VLT/MUSE and ALMA data from the PHANGS survey to derive SFR, molecular gas mass, and stellar mass surface densities across the galactic disks and to study the rSFMS ($\Sigma_\mathrm{SFR}$ versus $\Sigma_\mathrm{stellar}$), rKS ($\Sigma_\mathrm{SFR}$ versus $\Sigma_\mathrm{mol.gas}$), and rMGMS ($\Sigma_\mathrm{mol.gas}$ versus $\Sigma_\mathrm{stellar}$) relations in a sample of $18$ star-forming galaxies at spatial scales of $100$~pc, $500$~pc, and $1$~kpc. We tested for systematic differences induced by spatial scales considered, fitting approaches used, and assumptions made. Additionally, we have explored the $\Sigma_\mathrm{stellar + mol. gas} {-} \Sigma_\mathrm{SFR}$ correlation in our data to probe the effect of the mid-plane hydrostatic pressure of the disk as a regulator of the local SFR surface density. We applied a different approach to remove non-star-forming (diffuse) emission contaminating our SFR tracer before measuring the scaling relations, a different CO-to-H$_{2}$ conversion factor prescription, and a SFR tracer probing a longer timescale. Our main findings can be summarized as follows:

\begin{enumerate}
\item We have recovered all three scaling relations at a spatial scale of $100$~pc. Of the three relations, rMGMS shows the least scatter ($0.34$~dex) for our global data set, whereas the rKS is the relation that shows the highest level of consistency between different galaxies and across environments. Its higher scatter in the high resolution ($100$~pc) measurement is related to the inclusion of very low surface density SFR data points, following our methodology to recover all potential SFR emission. When probed at $1$~kpc scales, these data points are averaged over a larger region and the rKS shows the least scatter among the three relations (see Sec.~\ref{sec:global_results}).

\item At $100$ pc, we found that the scatter of the scaling relations follows  $\sigma_\mathrm{rSFMS} > \sigma_\mathrm{rKS} > \sigma_\mathrm{rMGMS}$. This is consistent with the expectation from the evolutionary scenario perspective, given $\tau_\mathrm{H\alpha} < \tau_\mathrm{CO} < \tau_\mathrm{stars}$, where $\tau$ corresponds to the duration each tracer is visible across the star formation cycle (see Secs.~\ref{sec:global_results} and ~\ref{sec:origin_scatter}).

\item We found significant variations in the studied scaling relations across different galactic environments. These variations are particularly strong in the case of the rSFMS and the rMGMS. The disk is the dominant feature in setting the slope, being the largest in area, while spiral arms share a similar slope, but are offset above the overall relations by up to $0.4$ dex. Bars lie systematically below the overall relations (see Sec.~\ref{sec:individual_gal}).

\item We searched for global parameters that could be driving the dispersion in the measured slopes between the galaxies in our sample. We found a correlation with $\Delta$MS, which implies that a global enhancement of SFR could change the slope of the scaling relations measured in a galaxy. However, we found a tighter correlation between the standard deviation of the distribution of the SFR tracer detection fraction with the deviation from the overall rSFMS slope. This can be explained by the fact that in galaxies with a  larger spread in the detection fraction distribution, outer regions have typically a lower detection fraction values. Thus, the low stellar mass surface density regime will be more affected by non-detections than the inner region with a higher detection fraction, resulting in a steeper rSFMS (see Secs.~\ref{sec:individual_gal} and ~\ref{sec:dispersion_slopes}). 

\item As long as non-detections are included in the measurement of the slope, the spatial scale of the data do not greatly or systematically impact the measured slope. The scatter on the other hand decreases at larger spatial scales (see Sec.~\ref{sec:spatial_scale_fiducial}).

\item Excluding non-detections from the analysis artificially flattens the relation at smaller spatial scales, resulting in a steepening when the analysis is carried out at larger spatial scales. This is because pixels with nonzero signal are averaged with the non-detection pixels at larger spatial scales. Furthermore, this effectively causes an artificial flattening of the rSFMS and the rMGMS at the low mass surface density end in the $100$~pc scale measurement (see Sec.~\ref{sec:slope_changes_ND}).

\item How much the slope of a galaxy is flattened when non-detections are excluded at $100$~pc spatial scale depends on the 2D distribution of ionized gas with respect to the stellar mass surface density (in the case of the rSFMS).  We computed the distribution of the H$\alpha$ detection fraction ($\mathrm{DF}_\mathrm{H\alpha}$) and we found a correlation between the mean of the detection fraction distribution and the change in slope when non-detections are not included as compared to the fiducial approach (see Sec.~\ref{sec:quantification_slope_change_ND}).

\item At all spatial scales, the scatter in the $\Sigma_\mathrm{b}$ versus $\Sigma_\mathrm{SFR}$ relation is higher than that seen for the rKS. 
We interpret this behavior such that the amount of available gas plays a primary role in locally regulating the SFR (see Sec.~\ref{sec:baryonic_vs_SFR}).

\item We removed all non-star-forming (diffuse) emission before resampling the SFR surface density maps, in order to reduce the level of contamination in our data. This strongly affects the level of emission at the edges of \hii\ regions and molecular clouds, and increases the scatter of the rSFMS and the rKS. This suggests that with our fiducial approach, we may underestimate the real scatter of these relations due to diffuse non-SF-related flux boosting the emission of the faintest SF pixels (see Sec.~\ref{sec:aperture}).

\item We have recomputed the rKS and the rMGMS under the assumption of a constant CO-to-H$_{2}$ conversion factor. Removing the metallicity-dependent radial gradient from the conversion factor leads to a slightly flatter rKS (${\sim}5\%$ flatter) and a slightly steeper rMGMS (${\sim}8\%$ steeper) (see Sec.~\ref{sec:conv_fact}).

\item We have recomputed the rSFMS and the rKS at $100$~pc scale using a longer timescale SFR tracer derived from the MUSE star formation histories. This longer timescale tracer smoothens the SFR values with respect to the shorter timescale tracer. Both relations are significantly flattened and their scatter is drastically reduced (see Sec.~\ref{sec:SFR_from_SFH}).
\end{enumerate}

Studying star formation in nearby galaxies at high physical resolution is a powerful tool to connect extragalactic observations with measurements in our own Galaxy. Assessing star formation scaling relations across different galaxy populations and quantifying systematic variations in different galactic environments will provide valuable insights in how galaxies grow and evolve in the local universe.

\begin{acknowledgements}
We thank the anonymous referee for their insightful 
comments which helped to improve the paper. This work was carried out as part of the PHANGS collaboration. 
DL, ES, HAP, TS, and TGW acknowledge funding from the European Research Council (ERC) under the European Union’s Horizon 2020 research and innovation programme (grant agreement No. 694343). 
FB acknowledges funding from the European Research Council (ERC) under the European Union’s Horizon 2020 research and innovation programme (grant agreement No.726384/Empire).
The work of AKL and JS is partially supported by the National Science Foundation (NSF) under Grants No.\ 1615105, 1615109, and 1653300.
E.C. acknowledges support from ANID project Basal AFB-170002.
JMDK and MC gratefully acknowledge funding from the Deutsche Forschungsgemeinschaft (DFG, German Research Foundation) through an Emmy Noether Research Group (grant number KR4801/1-1) and the DFG Sachbeihilfe (grant number KR4801/2-1). 
JMDK gratefully acknowledges funding from the European Research Council (ERC) under the European Union’s Horizon 2020 research and innovation program via the ERC Starting Grant MUSTANG (grant agreement number 714907).
KK gratefully acknowledges funding from the German Research Foundation (DFG) in the form of an Emmy Noether Research Group (grant number KR4598/2-1, PI Kreckel).
MQ acknowledges funding from the research project PID2019-106027GA-C44 from the Spanish Ministerio de Ciencia e Innovaci\'on.
AU acknowledges support from the Spanish funding grants AYA2016-79006-P (MINECO/FEDER), PGC2018-094671-B-I00 (MCIU/AEI/FEDER), and PID2019-108765GB-I00 (MICINN).
SCOG and RK acknowledge funding from the European Research Council via the ERC Synergy Grant ``ECOGAL -- Understanding our Galactic ecosystem: From the disk of the Milky Way to the formation sites of stars and planets'' (project ID 855130), from the DFG via the Collaborative Research Center (SFB 881, Project-ID 138713538) ``The Milky Way System'' (subprojects A1, B1, B2 and B8) and from the Heidelberg cluster of excellence (EXC 2181 - 390900948) ``STRUCTURES: A unifying approach to emergent phenomena in the physical world, mathematics, and complex data'', funded by the German Excellence Strategy.
Based on observations collected at the European Organisation for Astronomical Research in the Southern Hemisphere under ESO programmes IDs 094.C-0623, 098.C-0484 and 1100.B-0651. 
This paper also makes use of the following ALMA data: ADS/JAO.ALMA\#2013.1.01161.S, ADS/JAO.ALMA\#2015.1.00925.S, ADS/JAO.ALMA\#2015.1.00956.S and ADS/JAO.ALMA\#2017.1.00886.L. ALMA is a partnership of ESO (representing its member states), NSF (USA) and NINS (Japan), together with NRC (Canada), MOST and ASIAA (Taiwan), and KASI (Republic of Korea), in cooperation with the Republic of Chile. The Joint ALMA Observatory is operated by ESO, AUI/NRAO and NAOJ. The National Radio Astronomy Observatory is a facility of the National Science Foundation operated under cooperative agreement by Associated Universities, Inc.
\end{acknowledgements}

\bibliographystyle{aa}
\bibliography{biblio} 

\appendix
\section{Fitting of the $C_\hii$ parametrization}
\label{sec:ap_CHII}

\begin{figure}
    \centering
    \includegraphics[width=\columnwidth]{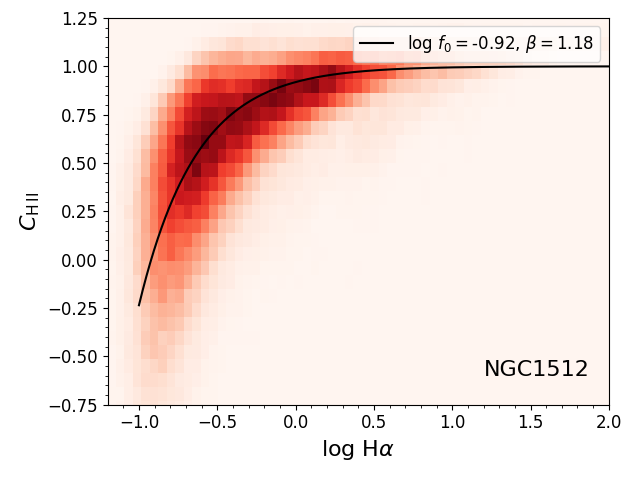}
    
    \caption{2D distribution of the $C_\hii$ fraction values of the pixels at the MUSE native resolution in NGC~1512 as defined in Eq.~\ref{eq:CHII-def}, as a function of the log H$\alpha$ flux surface density of each pixel, in units of $10^{-20}$ erg~s$^{-1}$ cm$^{-2}$~pc$^{2}$. The black solid line shows the best-fitting parametrization, as described in Eq.~\ref{eq:CHII}. The obtained $f_{0}$ and $\beta$ parameters are shown in the top of the panel.
    \label{fig:CHII_fit}}
\end{figure}

Figure~\ref{fig:CHII_fit} shows the $C_\hii$ values (i.e., the fraction of H$\alpha$ flux tracing star formation as defined in Eq.~\ref{eq:CHII-def}) of the pixels in the NGC~1512 map, as a function of their H$\alpha$ flux surface density. The solid black line shows the paramatrization defined in Eq.~\ref{eq:CHII}, and the best-fit parameters $f_{0}$ and $\beta$ are listed in the top of the panel. 

\section{${\Delta}\alpha_{\mathrm{overall}}$ as a function of {$\Delta$}MS for the rKS and the rMGMS}
\label{sec:ap_deltaMS_rKSrMGMS}
\begin{figure*}
    \centering
    \includegraphics[width = \textwidth]{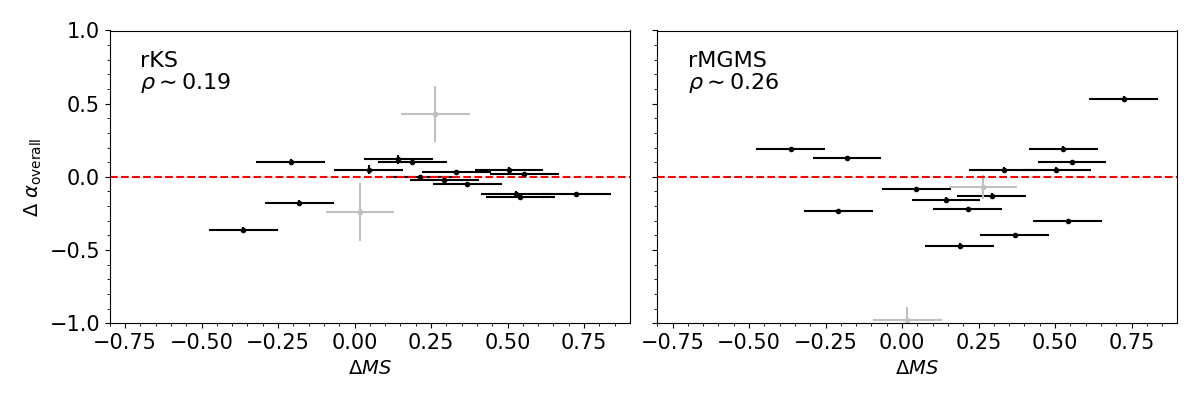}
    
    \caption{Differences in the slope measured for each individual galaxy with respect to the global measurements in Fig.~\ref{fig:global_all} for the rKS (left) and the rMGMS (right). Each dot represent a galaxy in our sample. The gray dots are NGC~2835 and NGC~5068, two low-mass galaxies. The PCC of the correlation is indicated in the top-left corner of each panel.}
    
    \label{fig:slope_var_vs_SFR_c}
\end{figure*}

Figure~\ref{fig:slope_var_vs_SFR_c} shows the dependence of ${\Delta}\alpha_{\mathrm{overall}}$ (as defined in Eq.~\ref{eq:global_def}) from {$\Delta$}MS, for the rKS and the rMGMS relations. We do not report any level of correlation between these parameters, and these figures are included here for completeness only.

\section{Measurement of the rKS and rMGMS at different spatial scales excluding non-detections}
\label{sec:ap_steepening}

\begin{figure*}
    \centering
    \includegraphics[width = \textwidth]{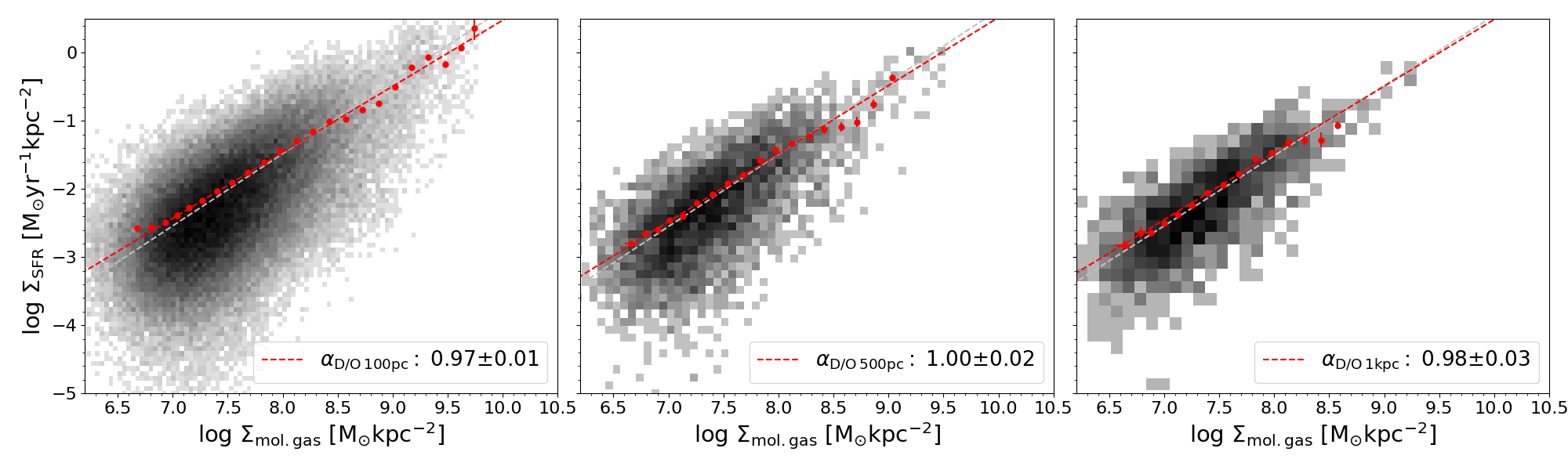}
    \caption{2D distribution of the resolved Kennicutt--Schmidt relation considering all galaxies in our sample at three spatial resolutions ($100$~pc, $500$~pc and $1$~kpc from left to right). Non-detection pixels have been excluded from the fit. The fiducial overall best-fitting power law at each spatial scale is marked with the gray dashed line for reference.
    \label{fig:global_KS_Hani}}
\end{figure*}

\begin{figure*}
    \centering
    \includegraphics[width = \textwidth]{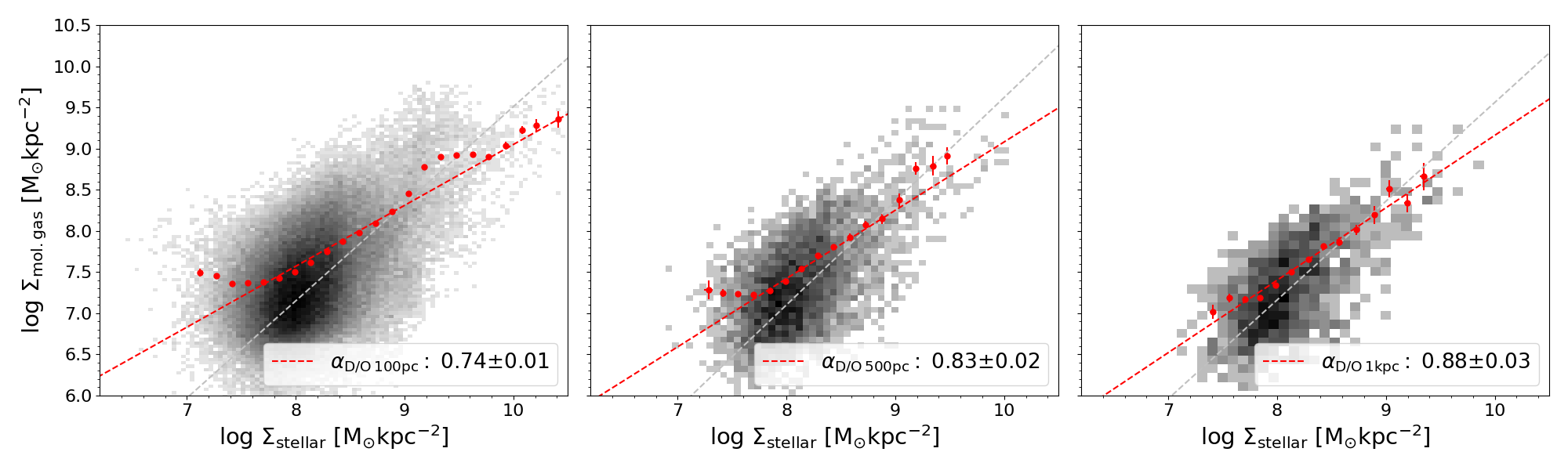}
    \caption{2D distribution of the molecular gas main sequence considering all galaxies in our sample at three spatial resolutions ($100$~pc, $500$~pc and $1$~kpc from left to right). Non-detection pixels have been excluded from the fit. This produces a systematic flattening at higher spatial resolution. The fiducial overall best-fitting power law at each spatial scale is marked with the gray dashed line for reference.
    \label{fig:global_MolGas_Hani}}
\end{figure*}

We include here the figures and tables with the measurement of the rKS and the rMGMS when non-detections are excluded from the analysis.
Figures~\ref{fig:global_KS_Hani} and \ref{fig:global_MolGas_Hani} show the obtained overall rKS and rMGMS respectively, at spatial scales from $100$ pc to $1$ kpc. Tables~\ref{tab:slopes_all_galaxies_KS_Hani} and ~\ref{tab:slopes_all_galaxies_MGMS_Hani} show the corresponding slope and scatter obtained for the same relations in each individual galaxy at the different spatial scales.
\begin{table*}
\centering
\begin{tabular}{ccccccc}
\hline
\hline
Target & $\alpha_{\mathrm{D/O}\,100\mathrm{pc}}$ & $\sigma_{\mathrm{D/O}\,100\mathrm{pc}}$ & $\alpha_{\mathrm{D/O}\,500\mathrm{pc}}$ & $\sigma_{\mathrm{D/O}\,500\mathrm{pc}}$ & $\alpha_{\mathrm{D/O}\,1\mathrm{kpc}}$ & $\sigma_{\mathrm{D/O}\,1\mathrm{kpc}}$ \\
\hline
NGC1087 & $1.06\pm0.02$ & $0.32$ & $1.04\pm0.02$ & $0.24$ & $1.00\pm0.02$ & $0.20$ \\
NGC1300 & $0.76\pm0.02$ & $0.39$ & $0.80\pm0.02$ & $0.28$ & $0.70\pm0.02$ & $0.24$ \\
NGC1365 & $0.92\pm0.01$ & $0.46$ & $0.87\pm0.01$ & $0.45$ & $0.89\pm0.01$ & $0.41$ \\
NGC1385 & $1.08\pm0.02$ & $0.32$ & $1.15\pm0.02$ & $0.26$ & $1.19\pm0.02$ & $0.20$ \\
NGC1433 & $0.54\pm0.02$ & $0.38$ & $0.76\pm0.02$ & $0.31$ & $0.70\pm0.02$ & $0.25$ \\
NGC1512 & $0.95\pm0.03$ & $0.37$ & $1.10\pm0.03$ & $0.28$ & $0.88\pm0.03$ & $0.24$ \\
NGC1566 & $0.93\pm0.02$ & $0.40$ & $0.96\pm0.02$ & $0.29$ & $0.86\pm0.02$ & $0.26$ \\
NGC1672 & $1.03\pm0.01$ & $0.42$ & $0.96\pm0.01$ & $0.37$ & $1.06\pm0.01$ & $0.33$ \\
NGC2835 & $1.45\pm0.20$ & $0.39$ & - & - & - & - \\
NGC3351 & $1.02\pm0.02$ & $0.36$ & $1.21\pm0.02$ & $0.26$ & $1.22\pm0.02$ & $0.09$ \\
NGC3627 & $1.06\pm0.02$ & $0.42$ & $1.05\pm0.02$ & $0.31$ & $1.15\pm0.02$ & $0.23$ \\
NGC4254 & $0.95\pm0.01$ & $0.37$ & $0.96\pm0.01$ & $0.28$ & $1.01\pm0.01$ & $0.23$ \\
NGC4303 & $0.86\pm0.01$ & $0.43$ & $0.81\pm0.01$ & $0.36$ & $0.79\pm0.01$ & $0.25$ \\
NGC4321 & $0.98\pm0.01$ & $0.40$ & $0.92\pm0.01$ & $0.33$ & $1.00\pm0.01$ & $0.26$ \\
NGC4535 & $0.98\pm0.03$ & $0.43$ & $1.01\pm0.03$ & $0.35$ & $0.92\pm0.03$ & $0.28$ \\
NGC5068 & $0.55\pm0.20$ & $0.48$ & - & - & - & - \\
NGC7496 & $0.82\pm0.02$ & $0.40$ & $1.10\pm0.02$ & $0.35$ & $1.00\pm0.02$ & $0.32$ \\
IC5332 & - & - & - & - & - & - \\
\hline
\end{tabular}
\caption{Slope ($\alpha$) and scatter ($\sigma$) for the rKS measured in each galaxy in our sample at spatial scales of $100$~pc, $500$~pc and $1$~kpc. The non-detections have been excluded from the slope measurement. }
\label{tab:slopes_all_galaxies_KS_Hani}
\end{table*}

\begin{table*}
\centering
\begin{tabular}{ccccccc}
\hline
\hline
Target & $\alpha_{\mathrm{D/O}\,100\mathrm{pc}}$ & $\sigma_{\mathrm{D/O}\,100\mathrm{pc}}$ & $\alpha_{\mathrm{D/O}\,500\mathrm{pc}}$ & $\sigma_{\mathrm{D/O}\,500\mathrm{pc}}$ & $\alpha_{\mathrm{D/O}\,1\mathrm{kpc}}$ & $\sigma_{\mathrm{D/O}\,1\mathrm{kpc}}$ \\
\hline
NGC1087 & $0.75\pm0.02$ & $0.27$ & $0.94\pm0.02$ & $0.20$ & $0.98\pm0.02$ & $0.18$ \\
NGC1300 & $0.73\pm0.01$ & $0.28$ & $0.87\pm0.01$ & $0.26$ & $0.89\pm0.01$ & $0.28$ \\
NGC1365 & $1.17\pm0.01$ & $0.38$ & $1.38\pm0.01$ & $0.37$ & $1.50\pm0.01$ & $0.29$ \\
NGC1385 & $0.68\pm0.02$ & $0.30$ & $0.87\pm0.02$ & $0.29$ & $0.91\pm0.02$ & $0.23$ \\
NGC1433 & $0.83\pm0.01$ & $0.23$ & $0.93\pm0.01$ & $0.20$ & $1.01\pm0.01$ & $0.20$ \\
NGC1512 & $0.55\pm0.01$ & $0.22$ & $0.75\pm0.01$ & $0.18$ & $0.87\pm0.01$ & $0.17$ \\
NGC1566 & $0.66\pm0.01$ & $0.30$ & $0.77\pm0.01$ & $0.27$ & $0.81\pm0.01$ & $0.23$ \\
NGC1672 & $1.04\pm0.01$ & $0.30$ & $1.09\pm0.01$ & $0.30$ & $1.19\pm0.01$ & $0.26$ \\
NGC2835 & $-0.02\pm0.06$ & $0.15$ & - & - & - & - \\
NGC3351 & $0.80\pm0.01$ & $0.20$ & $1.02\pm0.01$ & $0.15$ & $0.94\pm0.01$ & $0.11$ \\
NGC3627 & $0.52\pm0.02$ & $0.32$ & $0.56\pm0.02$ & $0.26$ & $0.58\pm0.02$ & $0.24$ \\
NGC4254 & $0.64\pm0.01$ & $0.29$ & $0.70\pm0.01$ & $0.23$ & $0.77\pm0.01$ & $0.16$ \\
NGC4303 & $0.69\pm0.01$ & $0.31$ & $0.75\pm0.01$ & $0.28$ & $0.77\pm0.01$ & $0.21$ \\
NGC4321 & $0.76\pm0.01$ & $0.28$ & $0.94\pm0.01$ & $0.24$ & $1.02\pm0.01$ & $0.21$ \\
NGC4535 & $0.85\pm0.02$ & $0.27$ & $0.96\pm0.02$ & $0.25$ & $1.04\pm0.02$ & $0.23$ \\
NGC5068 & $-0.20\pm0.05$ & $0.13$ & - & - & - & - \\
NGC7496 & $0.97\pm0.03$ & $0.28$ & $1.32\pm0.03$ & $0.23$ & $1.34\pm0.03$ & $0.16$ \\
IC5332 & - & - & - & - & - & - \\
\hline
\end{tabular}
\caption{Slope ($\alpha$) and scatter ($\sigma$) for the rMGMS measured in each galaxy in our sample at spatial scales of $100$~pc, $500$~pc and $1$~kpc. The non-detections have been excluded from the slope measurement. }
\label{tab:slopes_all_galaxies_MGMS_Hani}
\end{table*}

\section{Role of the detection fraction of the molecular gas tracer in the slope of the rMGMS}
\label{sec:ap_DF_rMGMS}

\begin{figure}
    \centering
    \includegraphics[width = \columnwidth]{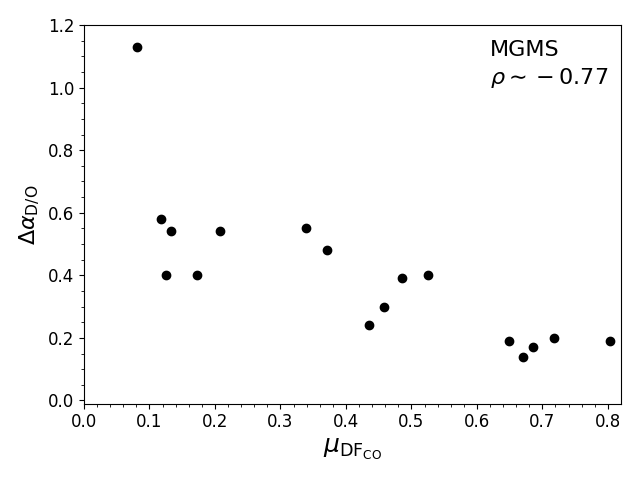}
    \caption{Difference between the slope of the rMGMS of each galaxy measured with the fiducial approach and the slope measured when non-detections are excluded, as a function of the mean detection fraction of the molecular gas tracer in each galaxy ($\mu_{\mathrm{DF}_{\mathrm{CO}}}$). All values are positive in the $y$-axis because excluding non-detections always leads to a flatter relation. The PCC of the correlation is indicated in the top-right corner of the panel.}
    \label{fig:CO_ff_vs_slope}
\end{figure}

\begin{figure}
    \centering
    \includegraphics[width = \columnwidth]{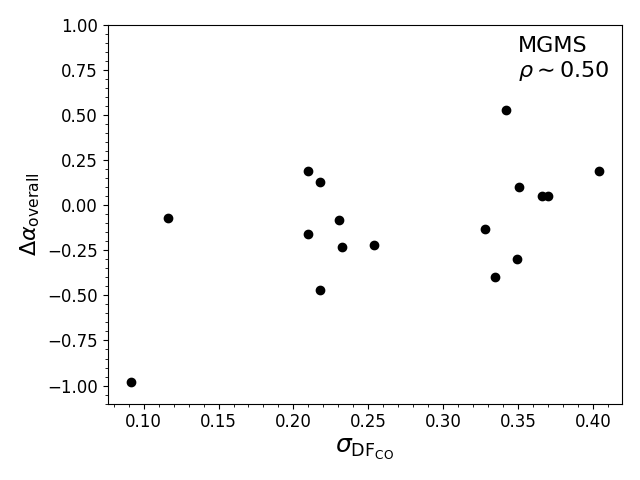}
    \caption{Difference between the slope of the rMGMS of each galaxy measured with the fiducial approach and the global measurement as a function of the standard deviation of the detection fraction distribution of the molecular gas tracer in each galaxy ($\sigma_{\mathrm{DF}_{\mathrm{CO}}}$). The PCC of the correlation is indicated in the top-right corner of the panel.}
    \label{fig:CO_ff_vs_slope_global}
\end{figure}

\begin{figure}
    \centering
    \includegraphics[width = \columnwidth]{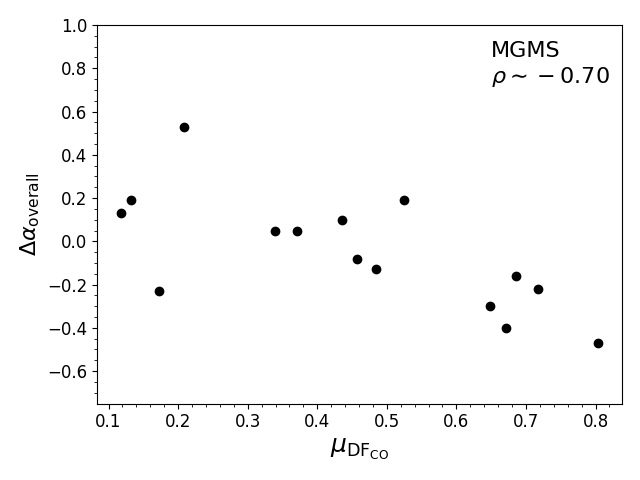}
    \caption{Difference between the slope of the rMGMS of each galaxy measured with the fiducial approach and the global measurement as a function of the mean of the detection fraction distribution of the molecular gas tracer in each galaxy ($\mu_{\mathrm{DF}_{\mathrm{CO}}}$). The PCC of the correlation is indicated in the top-right corner of the panel.}
    \label{fig:CO_ff_vs_slope_global2}
\end{figure}

Figure ~\ref{fig:CO_ff_vs_slope} shows how much the slope of the rMGMS in each galaxy varies when non-detections are excluded from its calculation, as a function of the mean in the distribution of the molecular gas tracer ($\mu_{\mathrm{DF}_\mathrm{CO}}$). We find a negative correlation (PCC $~\approx -0.77$), similar to that of the rSFMS. The outlier is the galaxy NGC~2835, which has a large uncertainty on its measured slope. Figure~\ref{fig:CO_ff_vs_slope_global} shows the difference between the slope of the rMGMS for each galaxy and the global measurement as a function of the standard deviation of the distribution of the molecular gas tracer ($\sigma_{\mathrm{DF}_\mathrm{CO}}$). The level of correlation we find here is modest (PCC$~\approx 0.50$), in contrast to that of the rSFMS slope differences and the SFR tracer distribution. This could be due to the completeness limit of our molecular gas tracer. As stated in Section~\ref{sec:global_results}, the rMGMS is the relation with the lowest total detection fraction (${\sim} 35\%$). The detection insensitivity to a fainter component should not impact our slope measurements, as our methodology is robust against the detection threshold. However, the detection fraction distribution will be affected. This leads to systematically lower values of $\mu_{\mathrm{DF}_\mathrm{CO}}$ for all galaxies. On the other hand, the impact in $\sigma_{\mathrm{DF}_\mathrm{CO}}$ is less systematic, and will depend on the mean of the distribution in each galaxy. We found a better correlation (PCC $\approx-0.70$) of $\Delta \alpha_\mathrm{overall}$ with $\mu_{\mathrm{DF}_\mathrm{CO}}$ instead. Figure~\ref{fig:CO_ff_vs_slope_global2} shows the difference in the rMGMS slope with the overall slope as a function of $\mu_{\mathrm{DF}_\mathrm{CO}}$, excluding the two galaxies with the more uncertain measurements (NGC~2835 and NGC~5068). The negative correlation implies that galaxies with a lower mean detection fraction of the molecular gas tracer have a steeper rMGMS.

\end{document}